\documentclass[lettersize,journal]{IEEEtran}
\usepackage{amsmath,amsfonts}
\usepackage{algorithmic}
\usepackage{algorithm}
\usepackage{array}
\usepackage[caption=false,font=footnotesize,labelfont=rm,textfont=rm]{subfig}
\usepackage{textcomp}
\usepackage{stfloats}
\usepackage{url}
\usepackage{verbatim}
\usepackage{graphicx}
\usepackage{cite}
\usepackage{booktabs}
\usepackage{tikz,xcolor}

\usepackage[implicit=false]{hyperref}

\hypersetup{hidelinks,
	colorlinks=true,
	allcolors=black,
	pdfstartview=Fit,
	breaklinks=true}

\definecolor{lime}{HTML}{A6CE39}
\DeclareRobustCommand{\orcidicon}{
\begin{tikzpicture}
\draw[lime, fill=lime] (0,0)
circle[radius=0.16]
node[white]{{\fontfamily{qag}\selectfont \tiny \.{I}D}};
\end{tikzpicture}
\hspace{-2mm}
}
\foreach \x in {A, ..., Z}{%
\expandafter\xdef\csname orcid\x\endcsname{\noexpand\href{https://orcid.org/\csname orcidauthor\x\endcsname}{\noexpand\orcidicon}}
}

\begin{document}

\title{Towards Covert APT Detection via Deviation-Aware Learning on Temporal Provenance Graph}

\author{Wenhan Jiang\hspace{-1.5mm}\orcidA{},
        Tingting Chai\hspace{-1.5mm}\orcidC{}, \IEEEmembership{Member,~IEEE},
        Kai Wang*\hspace{-1.5mm}\orcidB{}, \IEEEmembership{Member,~IEEE},   
        \\Hongke Zhang\hspace{-1.5mm}\orcidD{}, \IEEEmembership{Fellow,~IEEE}

\thanks{This work is supported by National Natural Science Foundation of China (NSFC) (grant number 62272129), Natural Science Foundation of Shandong Province (grant number ZR2023QF030) and Taishan Scholar Foundation of Shandong Province (grant number tsqn202408112).}
\thanks{Wenhan Jiang, Tingting Chai, and Kai Wang are with the School of Computer Science and Technology, Harbin Institute of Technology, Weihai, China, and are with Shandong Key Laboratory of Industrial Network Security, China (e-mail: 23S030142@stu.hit.edu.cn; ttchai@hit.edu.cn; liuhr@hit.edu.cn; dr.wangkai@hit.edu.cn).
}
\thanks{Hongke Zhang is with the School of Electronic and Information Engineering, Beijing Jiaotong University, Beijing 100044, China (e-mail: hkzhang@bjtu.edu.cn).
}
\thanks{
Kai Wang is the corresponding author.
}
}

\markboth{IEEE TRANSACTIONS ON NETWORKING}%
{Shell \MakeLowercase{\textit{et al.}}: A Sample Article Using IEEEtran.cls for IEEE Journals}

\IEEEpubid{}

\maketitle

\begin{abstract}
Advanced Persistent Threat (APT) have grown increasingly complex and concealed, posing formidable challenges to existing intrusion detection systems in identifying and mitigating these attacks. Recent studies have incorporated graph learning techniques to extract detailed information from provenance graphs, enabling the detection of attacks with greater granularity. Nevertheless, existing studies have largely overlooked the continuous yet subtle temporal variations in the structure of provenance graphs, which may correspond to surreptitious perturbation anomalies in ongoing APT attacks. Therefore, we introduce TFLAG, an advanced anomaly detection framework that for the first time integrates the structural dynamic extraction capabilities of temporal graph model with the anomaly delineation abilities of deviation networks to pinpoint covert attack activities in provenance graphs. This self-supervised integration framework leverages the graph model to extract neighbor interaction data under continuous temporal changes from historical benign behaviors within provenance graphs, while simultaneously utilizing deviation networks to accurately distinguish authentic attack activities from false positive deviations due to unexpected subtle perturbations. The experimental results indicate that, through a comprehensive design that utilizes both attribute and temporal information, it can accurately identify the time windows associated with APT attack behaviors without prior knowledge (e.g., labeled data samples), demonstrating superior accuracy compared to current state-of-the-art methods in differentiating between attack events and system false positive events. Our code and datasets are available at https://github.com/wangkai-tech23/TFLAG.
\end{abstract}


\section{Introduction}
\IEEEPARstart{I}{N} recent years, Advanced Persistent Threat (APT) have increasingly targeted government agencies \cite{ref1}, leading to a notable rise in both the frequency and severity of attacks affecting several critical sectors—including national defense, military operations, scientific research, education, and communications. In response to this growing threat landscape, intrusion detection technologies leveraging provenance graphs have emerged as effective tools.
A provenance graph \cite{ref2} represents a structured model derived from the core process level of traffic log architecture and encapsulates the operational behaviors and interactive data of the system. This includes multiple elements, such as processes, files, registry entries, network connections, and pipe transfers. Analyzing this complex data can accurately reflect the execution history and status transitions within the system, providing essential contextual insights and clues to trace and detect the origins of APT attacks.

Recent scholarly endeavors have significantly advanced the transformation of audit logs \cite{ref3} into provenance graphs, effectively redefining the challenge of APT detection as a multi-domain integration issue \cite{ref4,ref5,ref6,ref13,ref42,ref43}. The central issue in contemporary research is the effective utilization of extensive node and structural attribute information in provenance graphs to devise models capable of detecting APT attacks. The majority of approaches exploit the distribution of entities and events, as well as neighboring relationships within a provenance graph, to detect anomalous attacks \cite{ref4, ref5, ref6, ref7, ref8}. In detailed attack scenarios, these techniques extract structural information through nodal topological relationships, semantic properties, and the extensive distribution of edges, thereby facilitating the development of models capable of identifying aberrant structures. Given that APT attacks often mask their presence by reducing occurrence rates and blending seamlessly into normal system operations, reliance solely on structural data may not result in accurate detection.

Consequently, incorporating continuous timing data from crucial events into scenario analyses, and integrating temporal information from the provenance graph into graph learning models, has emerged as a promising new direction \cite{ref9, ref10, ref23}, enabling more accurate detection of APT attacks based on event sequences. However, existing studies frequently neglect the role of continuous yet subtle temporal factors and the distribution of anomalous behavior in graph representation learning. Relying exclusively on discrete spatiotemporal information between nodes and events complicates the detection of subtle malicious activities buried in extensive normal logs and hampers the accurate differentiation between false positives that depart from typical behavior and genuine attacks.

We introduce the TFLAG framework, a self-supervised approach designed to address the aforementioned challenges in APT detection by leveraging dynamic provenance graph data from benign scenarios to identify unknown, concealed, and complex fine-grained attack behaviors. This novel anomaly detection framework that combines the structural dynamic extraction capabilities of temporal graph models with the anomaly detection abilities of deviation networks \cite{ref14}. This integration allows for more comprehensive extraction of contextual information from temporal provenance graphs, addressing the limitations of previous models in capturing subtle temporal variations. To address the scarcity of labels in real-world scenarios and reduce false positives from benign events, the TFLAG framework utilizes temporal graph models to aggregate features from historical neighbors and continuously self-adjusts its learning by predicting event types. Additionally, it enhances the structural distinctions between anomalous and normal events by recording anomaly scores and constructing a deviation network. In the detection phase, this cohesive learning model proficiently identifies high-scoring anomalous events and effectively discriminates between attack events and false positives through deviation scores. 

Our contributions are summarized as follows:
\begin{itemize}
\item We propose a self-supervised provenance graph attack detection framework designed to process logs from auditing systems into dynamic temporal graphs, effectively capturing the continuous yet subtle temporal variations in the structure of provenance graphs to detect attacks. To address real-world scenarios where attack labels are scarce or unavailable, this framework learns the temporal graph structure under benign conditions without requiring labeled data.

\item We introduce an innovative method that integrates temporal sequence models with deviation networks, marking its first application to the detection of APT attacks in provenance graphs. By embedding temporal features into the graph structure and linking these to a deviation network derived from the anomaly scores of historical events, this approach extracts the continuous yet subtle temporal variations in the structure of provenance graphs, enabling more accurate detection of covert perturbation anomalies in APT attacks.

\item We evaluated the TFLAG framework on widely utilized APT datasets, conducting both window-level and graph-level anomaly detection evaluations. The results demonstrate that TFLAG outperforms existing state-of-the-art methods in accurately detecting APT attacks. Furthermore, the findings underscore the superiority of our approach over current state-of-the-art methods.
\end{itemize}

The structure of this paper is organized as follows: Section II reviews relevant work within the current research landscape. Section III introduces the preliminaries and motivation. Section IV outlines the threat model of our system. Section V describes the overarching framework of the TFLAG system. Section VI details the experimental methodology, and Section VII presents our conclusions.

\section{Related Work}
We have investigated the detection of APT attacks in provenance graphs through the application of graph neural networks. Consequently, our focus has shifted toward a review of the relevant literature that discusses these aspects.

\subsection{APT Detection on Static Provenance Graph}

Contemporary research predominantly explores complex structural issues by integrating graph methodologies with APT detection strategies. Nonetheless, some studies continue to employ traditional and statistical methods to detect anomalies within system entities \cite{ref6,ref11,ref12}. These approaches utilize causality analysis or sequential pattern learning to extract and identify relationships and features effectively at a superficial level, thereby yielding reliable results in detecting abnormal nodes and sequences, as well as in measuring the distances and scores between anomalous and normal behaviors. However, the reliance on superficial data associations limits the ability to uncover deeper harmless structures amidst the complex, evolving scenarios of APT attacks, ultimately undermining the stability of detection performance over prolonged periods. 

To mitigate this limitation, recent advancements have incorporated sophisticated graph learning techniques into APT anomaly detection \cite{ref4, ref5, ref7, ref8, ref9, ref10}, offering a more versatile and effective solution. For instance, early applications such as ThreaTrace \cite{ref5} have utilized the GraphSAGE \cite{ref46} model for node-level anomaly detection, focusing primarily on understanding neighborhood structures and identifying anomalies through these behavioral patterns. R-CAID \cite{ref4} has introduced root node embedding methods, connecting each graph node back to its foundational root cause and employing graph attention mechanisms for dynamic classification. This method proficiently captures the structural nuances of trace graphs while minimizing disturbances from unrelated nodes. However, these systems generally overlook the impact of temporal dynamics on graph structures. MAGIC \cite{ref7} employs a graph autoencoder (GraphMAE \cite{ref19}) to learn the graph's characteristics by masking benign nodes, with joint oversight from masked feature reconstruction and sample-based structural reconstruction to detect anomalies through node outliers. Nevertheless, MAGIC does not consider potential alterations in graph structure due to extended network traffic over time. Although Flash \cite{ref8} takes into account the spatial and temporal contextual semantics in data modeling, its semantic analysis is performed prior to model training, which results in temporally discrete data. This could result in overlooked significant APT attack features demanding continuous temporal relevance, leading to potentially incomplete data capture during model training.

\subsection{APT Detection on Dynamic Provenance Graph}

Recent studies have integrated continuous time into traceability-based intrusion detection systems. Among these, Unicorn \cite{ref9} epitomizes an initiative by introducing flow graphs that transform provenance graphs into incrementally updatable sketches, capturing distinct temporal snapshots of the system. This approach enables Unicorn to monitor current vectors and compare them against historical sketches of typical behavior to detect anomalies. However, Unicorn faces challenges in detecting rare, subtle anomalies obscured within numerous sketches of normal activities. To extract historical correlation information more efficiently from provenance graphs, TBDetector \cite{ref48} employs a Transformer-based method to derive feature sequences from the transformed provenance graphs. The anomaly score of the sequence is calculated using both similarity scores and isolation scores. While this approach resembles Unicorn's method of converting data into flow graphs, it similarly lacks effective use of temporal attributes and interpretability of results. In contrast, JBEIL \cite{ref47} employs an inductive learning strategy using sequence diagrams. It processes log attributes through graph mapping to delineate the structure of these diagrams, focusing on the static phases of lateral movements in APT attacks. Regrettably, detection modules tailored for specific attack phases may not be as effective when applied to complex, real-world scenarios. Meanwhile, Kairos \cite{ref10} harnesses the TGN \cite{ref18} model, lauded for its effectiveness in dynamic graphs, to advance provenance graph anomaly detection. It aggregates dynamic node features through a memory module while identifying anomalies using reconstructed errors. However, Kairos is hindered by its inability to effectively distinguish specific low-frequency APT attacks hidden within a large volume of benign data and occasional system behaviors, thus increasing the likelihood of misclassifying benign edges as anomalous.

\begin{figure*}[!t]
\centering
\includegraphics[width=0.9\textwidth]{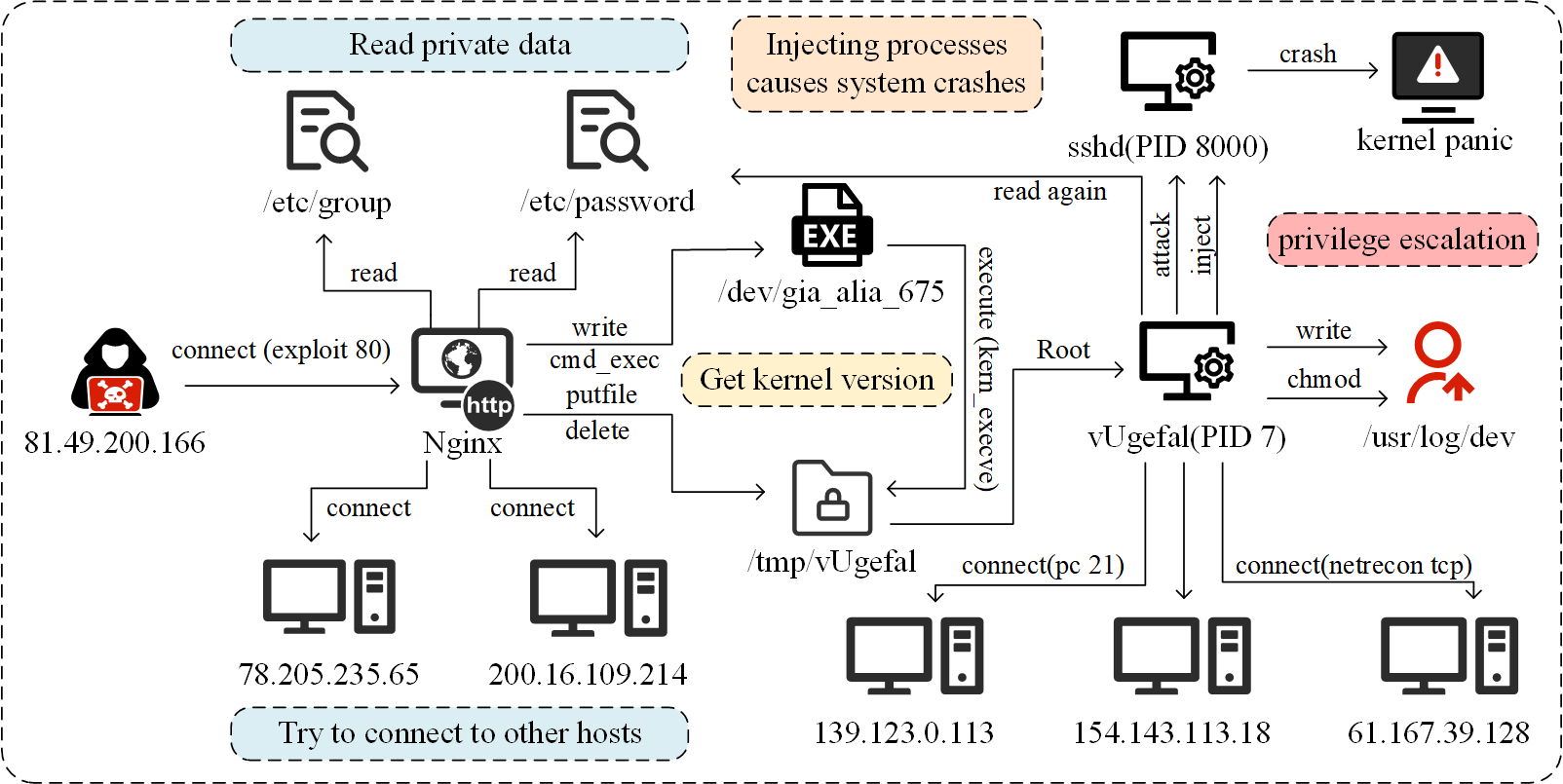} 
\caption{The DARPA E3-CADETS scenario features a comprehensive attack provenance graph that encompasses various entity and event types. The attacker launched multiple assaults on FreeBSD using Nginx and secured administrative backend privileges during the second attempt.}
\label{fig_1}
\vspace{-1.0em}
\end{figure*}

\section{Preliminary \& Motivation}
This section introduces the definitions and methods associated with temporal provenance graphs, followed by a description of how TFLAG addresses the limitations of current APT detection systems that are based on provenance graphs. This is demonstrated through a real attack scenario in the CADETS (Darpa E3) dataset.

\subsection{Temporal Provenance Graph}

Most traceability data comprise attack events that unfold over time, rather than merely focusing on the relationships delineated through graphical structures. Overlooking the impact of the temporal dimension on the connectivity features between nodes can hinder the effective tracking of continuous attack behaviors. As the system progresses, the conditions of nodes, the attributes of events, and their interconnections evolve dynamically. It is therefore essential to identify and differentiate boundary anomalies prompted by such changes. To encapsulate these characteristics, we model the dynamic timeline \( G = (V, E) \) as a process. Here, \( V = \{p_1, p_2, p_3, \dots\} \) denotes our process entity nodes, and \( E = \{e(t_1), e(t_2), \dots, e(t_m)\} \) represents a stream of dynamic connection events. Each event \( e(t) = (p_1, p_2, t, x_{p_1 p_2}) \) signifies the interaction that took place between process \(p_1\) and process \(p_2\) at time \( t \), melding the features of the two nodes to form edge characteristics. Due to the continuity and synchronization inherent in time-dynamic graphs, the provenance graph may exhibit multiple connection types, each generating specific interaction patterns influenced by the topology of the graph.

In real scenarios, data collection tools used in system monitoring generate substantial amounts of traffic data rapidly, wherein APT attacks are frequently hidden using sophisticated methods. The abundant information inherent in continuously occurring events challenges the effectiveness of coarse-grained, discrete snapshot-based data analysis. The system must analyze numerous time-segmented windows to differentiate between anomalous and normal labels for supervised training. Moreover, it struggles to identify the subtle anomaly patterns of APT attacks amidst ongoing temporal variations. To address these issues, TFLAG employs a self-supervised approach that eschews reliance on labeled anomalies. It integrates features from both source and target nodes at each timestep within the temporal flow \( e(t_i) \) of graph \( G \) , to predict the type of each edge connection. Concurrently, it logs accurate prediction labels to build a deviation network, iteratively optimizing the model as new data emerges to minimize reconstruction errors, thereby effectively distinguishing between normal and attack events in the future.

\subsection{Motivation Example}

In the CADETS attack scenario, adversaries exploit middleware vulnerabilities to initially access and control the target server, aiming to retrieve sensitive files to facilitate further privilege escalation. These incursions often mimic legitimate user activities in system logs, thereby significantly complicating the detection of attacks in a timely manner. Should an error occur during an attack phase, adversaries refrain from restarting immediately; instead, they adopt a strategy of low-frequency traffic to evade detection systems that monitor abnormally high-frequency activities. Upon securing a stable connection, attackers maintain prolonged latency to consolidate control, which prepares the groundwork for more severe future attacks.

In this attack sequence, detailed in Fig.~\ref{fig_1}, the adversary targets the \texttt{Nginx} vulnerability on a \texttt{FreeBSD} system. Throughout this operation, the attacker utilizes classic APT strategies, exploiting vulnerabilities for lateral movement and gradually increasing control within the target environment. This scenario underscores the critical challenges in network and system defense, particularly the need for differentiating normal from anomalous activities and detecting adversarial actions early in multi-step attacks.

We conduct a comprehensive analysis of the aforementioned scenario and address the issues identified in the previous studies as referenced in Section II. By effectively leveraging the advantages of the TGAT model \cite{ref15}, TFLAG dynamically captures information about sequential neighbor events occurring around nodes and aggregates even minor perturbations within normal behavior into features.
In contrast to TGN, which retains historical node feature information through a memory module, TGAT aggregates the temporal features of neighboring nodes and the node in focus at the current moment. This approach provides an advantage in capturing sudden interactions between network nodes (\texttt{Nginx} connect) and file access (\texttt{vUgefal} read) events.
During the initial stages of an attack, this scenario generates a substantial volume of IP interactions, encompassing both normal system behaviors and the actions of the attacker. Consequently, if reconstruction is the only factor considered in detection, it becomes challenging to differentiate between system false positives and real attacks. To more accurately distinguish anomalies from normal events and reduce false alarms, TFLAG not only processes sequential events but also stores the anomaly scores of aggregated features in a memory module for further fine-grained differentiation. As it reconstructs node features, TFLAG dynamically forms a deviation network from the anomaly scores of events over time, thereby increasingly distancing the attack behaviors from the network structures of normal events. This method not only considers the impact of temporal information on APT attack detection but also enhances the precise identification of attack events.

\begin{figure*}[!t]
\centering
\includegraphics[width=0.9\textwidth]{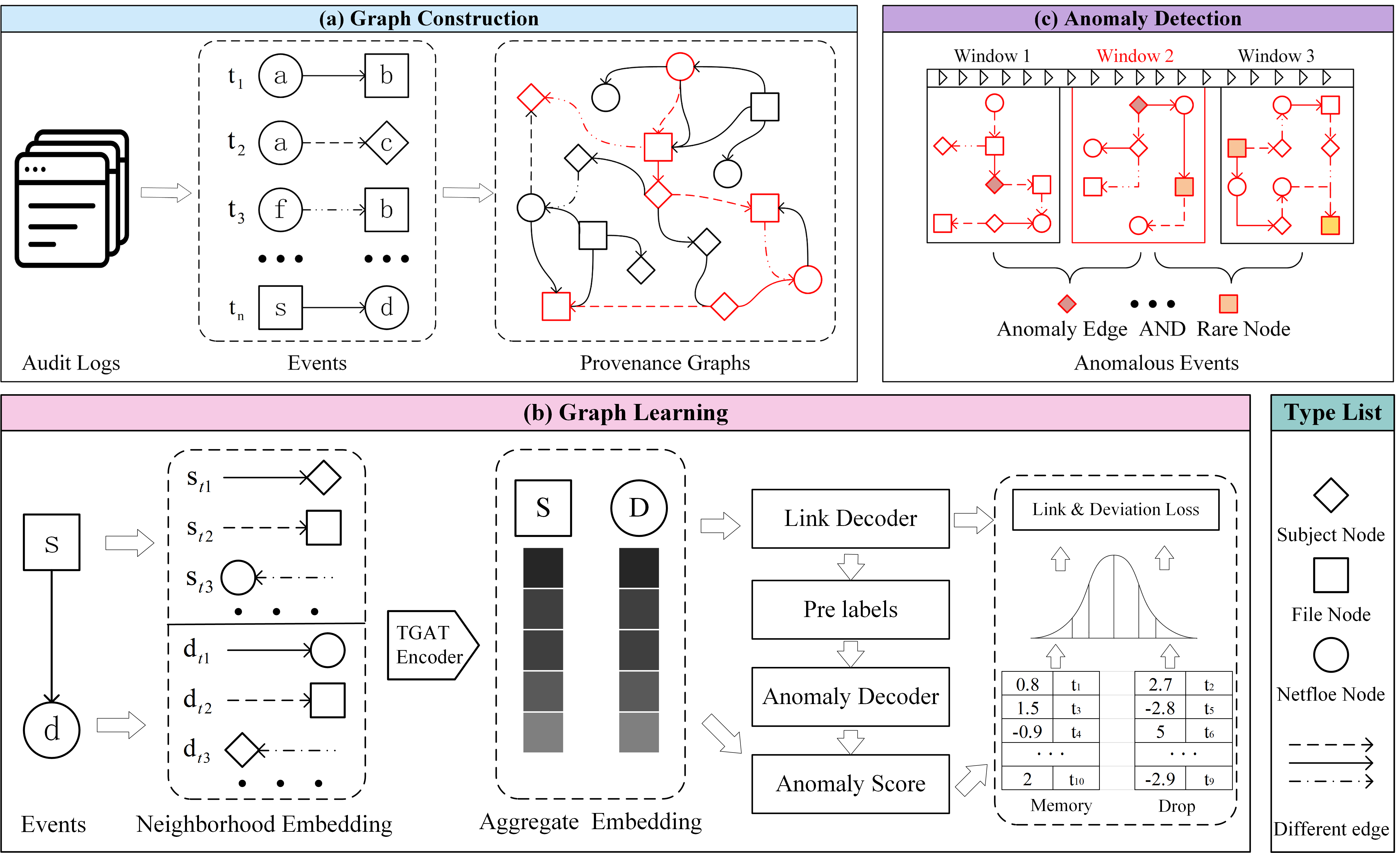} 
\caption{The architecture of the TFLAG framework comprises three integral modules: (a) Process-Level Provenance Graph Constructor, (b) Abnormal Behavior Learning Module Based On Dynamic Graphs and Deviation Networks, and (c) Abnormal Event Window Detector.}
\label{fig_2}
\vspace{-1.0em}
\end{figure*}

\section{Threat Model}

This section outlines system-specific threat models and their corresponding assumptions, establishing a robust theoretical foundation for developing effective threat detection methodologies. The model takes into account potential attacker behaviors and critical system characteristics, specifically tailored to address detection challenges posed by contemporary APT scenarios.

\subsection{Threat Model Assumptions}
We contend that an attacker's primary objective is to gain control over the system while maintaining a persistent presence. This objective can be achieved through exploiting software vulnerabilities, implanting backdoors, or employing other common attack vectors. To enhance the stealthiness of their operations, attackers often intertwine malicious activities with legitimate system processes, making these malicious behaviors difficult to detect with traditional rule- or signature-based methods. Furthermore, the use of zero-day exploits introduces entirely new behavioral patterns, posing significant challenges for self-supervised anomaly detection systems.
\subsection{System Environment Assumptions}
Prior to integrating data into the anomaly detection framework, it is assumed that both the behavioral system and logging software operate within a secure and reliable environment \cite{ref20,ref21}. The implementation of tamper-proof storage technologies ensures the integrity and accuracy of all audit log data, crucial for detecting and analyzing malicious activities. Additionally, we assume the system can comprehensively observe normal behavior during the initial learning phase, thereby establishing a baseline model. By preserving log integrity \cite{ref22}, our system is equipped to trace the origins of malicious actions, reconstruct the attack trajectory, and provide substantial evidence to support subsequent security responses.

\begin{figure}[!t]
\centering
\includegraphics[width=2.5in]{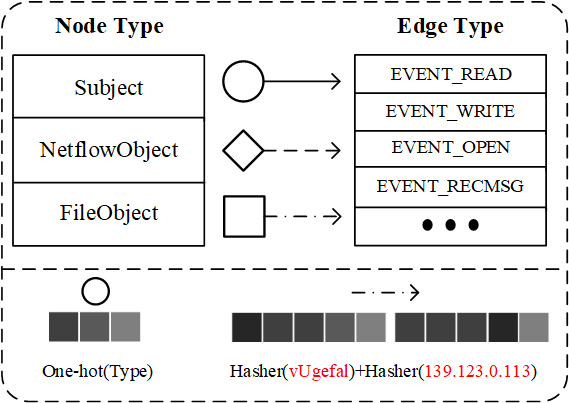}
\caption{The TFLAG encoding methodology for attributes of node and edge categories.}
\label{fig_3}
\vspace{-1.5em}
\end{figure}

\section{TFLAG Framework}

TFLAG is an anomaly-based system designed to detect stealthy APT attack behaviors. It operates without requiring any prior knowledge of attack characteristics, relying solely on the structure and attributes of the dynamic provenance graph to identify anomalous activities. By integrating the graph model with a deviation network, the system effectively identifies subtle disturbances induced by attacks, thereby efficiently differentiating between genuine threats and system false positives. Fig.~\ref{fig_2} provides an overview of the TFLAG framework.

In Module (a), continuously occurring real-time events are sequentially transformed into a temporal provenance graph with inherent attributes. This graph dynamically updates its structure with each new event, ensuring that the temporal and attribute information of each event is comprehensively captured, thus avoiding the segmentation of the overall structure into discrete fragments.

In Module (b), historical aggregation vectors for nodes at each endpoint are generated by merging all prior neighbor relationship vectors, including adjacent nodes and edges, to capture the subtle and continuous temporal variations within the structure of the provenance graph. Leveraging the learned node features, edge relationships are reconstructed, and the reconstruction error for each edge is calculated. Additionally, a repository is established to dynamically record the overall statistical distribution of predicted correct features, serving as a reference for computing deviation loss. The model undergoes self-supervised optimization using a combination of link prediction loss and deviation loss, eliminating the need for attack labels.

In Module (c), the reconstruction loss and deviation loss are sequentially integrated within a specific time window, allowing TFLAG to accurately identify edges with the highest abnormal values during each anomalous period. As events deviating from normal system behavior can also result in elevated anomaly scores, false attack windows may appear within the candidate time window queue. To effectively discern the true attack windows, we combine the reconstruction loss anomaly score, deviation loss, and node rarity, utilizing these integrated metrics to identify authentic attack windows marked by significant outliers (e.g., window 2 in Module (c)). Building on this, TFLAG recursively evaluates adjacent anomalous windows preceding and succeeding the central attack window to ascertain whether they represent false positives, ultimately enabling a comprehensive tracing of the complete attack sequence.

\subsection{Provenance Graph Construction}
The TFLAG system incrementally constructs a provenance graph by processing streaming system logs in batches. Echoing methodologies from previous research on provenance graph construction, the TFLAG system selectively processes logs from various systems, emphasizing objects and event types pertinent to APT detection. This approach effectively minimizes the influence of extraneous data on model training and organizes events chronologically.

The development of the provenance graph centers on three principal node types: \textbf{Subject}, \textbf{NetflowObject} and \textbf{FileObject}, which foster diverse event connections. To minimize storage requirements while preserving detailed log information, we prioritize processing node data by extracting its type and relevant process details. Specifically, the Subject node is linked to command line or execution details, the File node corresponds to the file path, and the Netflow node associates with IP addresses and port numbers. For edges, we derive the node IDs, their types, and the timestamp of each event. This integration of varied node and process information encapsulates the occurrence of events, with node attributes incorporating extensive logs and foundational data. In processing edges, we adopted a unified encoding of source and destination nodes based on their characteristics.

Fig.~\ref{fig_3} illustrates node features using one-hot encoding to depict the subject types at either end of an event. Traditional approaches to representing rich system process attributes typically result in vectors that are highly cardinal and sparse, reducing training efficiency. To overcome these challenges, we apply hierarchical hashing techniques \cite{ref23} to reduce vector sparsity by mapping high-cardinality categorical features into a fixed-dimensional vector space. The hashing function translates input features into hash values, which are then used to embed features into a fixed-dimensional vector based on these values. 

\subsection{Graph Encoder}\label{encoder}

In this model, the encoder is designed to tackle a challenge not yet addressed by previous research: identifying subtle and continuous anomalies that evolve dynamically within extensive datasets, thereby ensuring comprehensive coverage of all aspects of an APT attack process. Traditional APT detection methods have been ineffective in this regard because they predominantly depend on discrete time segments and often neglect the valuable information contained in historical data. To tackle the challenge of accurately identifying continuous yet subtle anomalous changes, this paper introduces the TGAT \cite{ref15} into the domain of APT attack detection for the first time, facilitating the encoding of nuanced anomalous changes and providing a robust framework to ensure comprehensive and complete detection. This approach allows for the aggregation of nodes, edges, and events through an attention mechanism \cite{ref45}. The encoder comprises multiple layers of Graph Neural Networks (GNNs) that process graph \( G \) at each time point  \( t \), extracting the representation \( z_i(t) \) of all nodes via message-passing mechanisms.

TGAT addresses the dispersed, phase-based nature of attacks more effectively than continuous-time models like TGN. APT attacks are characterized by intermittent, abrupt events and sudden structural changes in network graphs, such as unexpected new connections or anomalous node interactions. While TGN continuously updates embeddings for long-term dependencies, TGAT's temporal aggregation approach better captures the sporadic bursts and phase-based behaviors typical of APT campaigns, making it superior at detecting the short-term changes that continuous models may miss.

At the \( k \)-th layer, the TGAT framework updates node representations as follows: \textbf{Aggregation}. The representation of neighbor node \( v_j \), combined with edge attributes \( x_{ij} \), temporal encoding \( \phi(\Delta t) \), and the attention weight \( \alpha_{i j}(t) \), is aggregated as shown below: 
\begin{equation}
\label{ex1}
\sum_{v_{j} \in \mathcal{N}\left(v_{i}, t\right)} \alpha_{i j}(t) \cdot \mathbf{h}_{j}^{(k-1)}(t) \cdot \phi\left(\Delta t_{i j}\right) \cdot x_{i j}.
\end{equation}

\textbf{Combination}. The representation at the current layer $ h_i^{(k)}(t) $ for node $ v_i $ is derived by merging its previous layer's representation $ h_i^{(k-1)}(t) $ with the aggregated data from its neighbors. This process preserves the node’s historical continuity and incorporates updates from its adjacent nodes. This step ensures that the node representations reflect temporal changes in their local graph structure, as illustrated below: 
\begin{equation}
\begin{aligned}
h_i^{(k)}(t) &= \text{COMBINE} \Bigg( h_i^{(k-1)}(t), \\
&\quad \sum_{v_j \in \mathcal{N}(v_i;t)} \alpha_{ij}(t) \cdot h_j^{(k-1)}(t) \cdot \phi(\Delta t_{ij}) \cdot \mathbf{x}_{ij} \Bigg).
\end{aligned}
\end{equation}

Key components include $ \alpha_{ij}(t) $, the attention mechanism weight \cite{ref25}, and $ \phi(\Delta t_{ij}) $, the time encoding function that transforms time differences into a continuous feature representation.

Following the \( k \)-th layer, as shown in Module (b), the final embedding representation  \(\mathbb{z}_{i}(t)=\mathbf{h}_{i}^{(K)}(t)\) for nodes at both ends of an event can be reliably obtained. In the graph encoding process, the aforementioned framework enhances dynamic graph modeling capabilities through the utilization of temporal encoding and deep feature aggregation.

\begin{figure}
\centering
\includegraphics[width=2.7in]{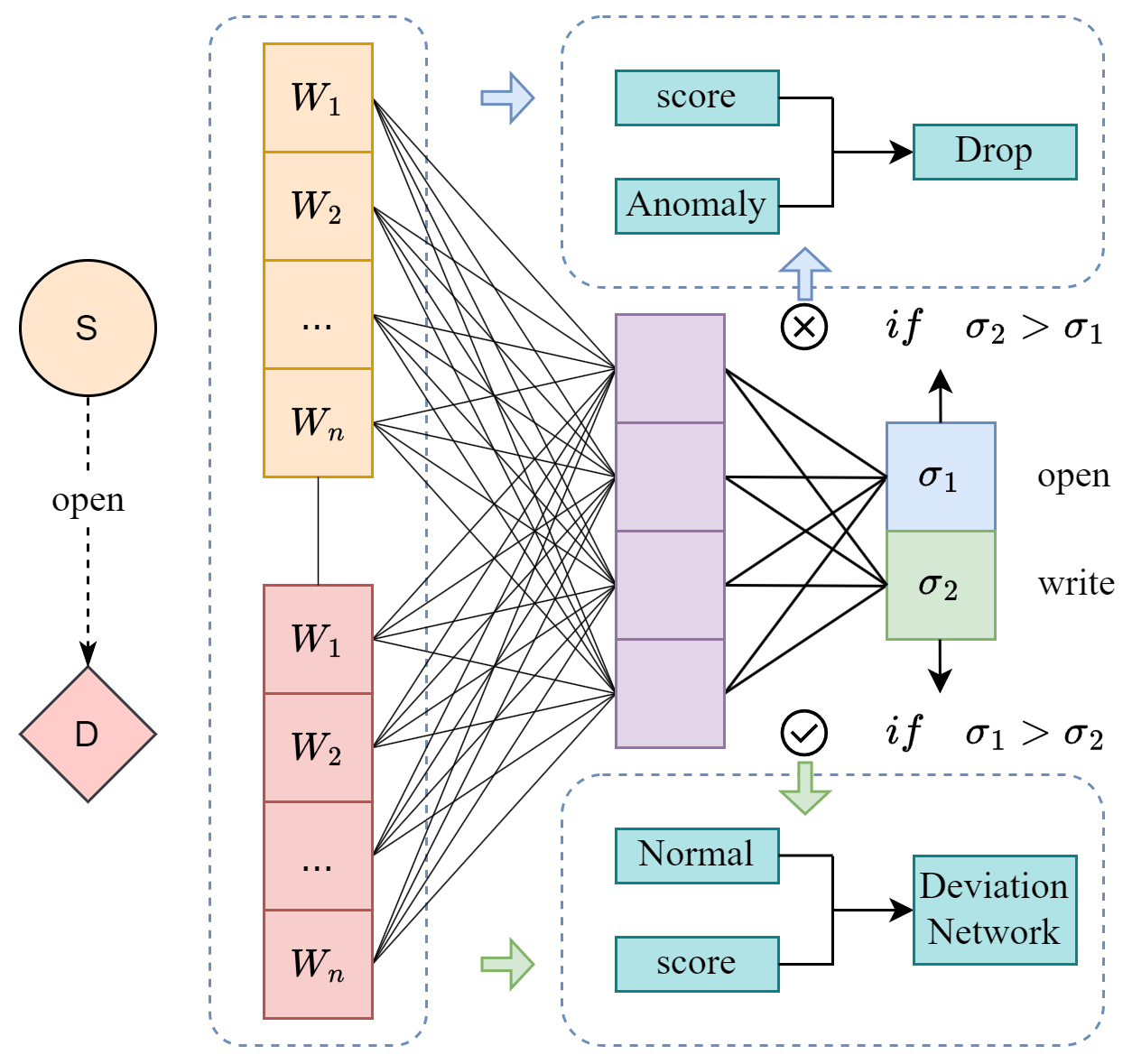}
\caption{The decoding process involves aggregated source and destination node features, under the assumption that there are only two edge types: \textbf{write} and \textbf{open}}
\label{decoder}
\vspace{-1.0em}
\end{figure}

\subsection{Graph Decoder}

Section V-B addresses encoding completeness, but recent research reveals another challenge: distinguishing between subtle anomalous changes indicating actual APT attacks versus sporadic system anomalies. This contributes to substantial false positive rates in anomaly-based detection methods. TFLAG addresses this by integrating event type prediction with deviation analysis in network anomaly score distribution through an innovative decoder structure.

As illustrated in Fig.~\ref{decoder}, TFLAG utilizing the context aggregation features derived from the encoder, a multilayer perceptron (MLP) \cite{ref26} is developed as a decoder to predict edge types. This decoder processes the aggregated features \(z\) through a series of nonlinear transformations, culminating in a probability distribution \(Q\left(e_{t}\right)\) indicative of the edge type: 

\begin{equation}
\label{ex3}
Q\left(e_{t}\right)=\operatorname{Softmax}\left(g\left(z ; \theta_{g}\right)\right).
\end{equation}

The function \(g\left(z ; \theta_{g}\right)\) denotes the nonlinear transformation with parameters  \(\theta_{g}\), and the probability distribution \(Q\left(e_{t}\right)\) reflects the confidence level for each possible edge type. During training, the loss function \(L_{\text {pred }}\) is minimized between the predicted probability distribution \(Q\left(e_{t}\right)\) and the actual edge type \(O\left(e_{t}\right)\) to enhance the accuracy of edge type predictions:

\begin{equation}
\label{ex4}
L_{\text {pred }}=\mathcal{D}\left(Q\left(e_{t}\right), O\left(e_{t}\right)\right).
\end{equation}

Here, \(\mathcal{D}\) represents the loss function measuring the discrepancy between the predicted and actual distributions. To further enhance anomaly detection, we introduce an anomaly scoring mechanism based on edge type prediction. For each sample that accurately predicts an edge type, we concatenate the embedding representations \(z_{\mathrm{src}}\) and \(z_{\mathrm{dst}}\) of the source and target nodes. These joint features are then input into the anomaly scoring module, which computes the anomaly score for the joint feature through the parameterized function \(h\left(\left[z_{\mathrm{src}}, z_{\mathrm{dst}}\right] ; \theta_{h}\right)\): 
\begin{equation}
\label{ex5}
A_{i}(t)=h\left(\left[z_{\text {src}}, z_{\text {dst}}\right] ; \theta_{h}\right).
\end{equation}

In provenance graphs, normal events typically exhibit consistent anomaly score distributions \(A_{i}(t)\), while attack events show significant deviations. Traditional systems relying solely on anomaly scores or reconstruction losses produce high false positive rates. TFLAG's novel decoder generates dynamic labels by predicting event categories and constructs a deviation network that continuously records normal event anomaly score distributions during learning. This realigns deviating scores within the established normal distribution range, enhancing system robustness. The decoder periodically recalibrates the overall anomaly score distribution to maintain accuracy for recent abnormal events, preventing outdated scores from distorting evaluations. Through dynamic event prediction and dual deviation network optimization, TFLAG accurately detects APT attacks while effectively minimizing false positives.

\subsection{Learning Procedure}
The optimization of the entire network utilizes cross-entropy loss \(L_{\text {pred }}\) and deviation loss \(L_{\text {dev}}\) \cite{ref27,ref28} concurrently. The cross-entropy loss, associated with the prediction task \cite{ref29}, measures the discrepancy between the predicted and actual probability distributions of the edges. By iteratively minimizing this loss, the model's edge reconstruction and classification accuracy are progressively improved.

Deviation networks significantly enhance anomaly detection across various domains. However, in dynamic graphs, the deviation loss is significantly influenced by temporal factors. Over time, different events exhibit distinct abnormal variations, and using a standard normal distribution as a reference score complicates handling diverse scenarios. The model relies on an adjustable repository to generate reference scores, assuming that the distribution of the initial anomaly scores follows a Gaussian distribution—an effective approach for various datasets \cite{ref29}. Additionally, incorporating time labels introduces an attenuation effect, thereby differentially weighting statistical samples across varying time spans. With these timestamped reference scores, deviations are calculated as follows:

\begin{equation}
\label{ex6}
\operatorname{dev}(\mathbf{x}, t)=\frac{\phi(\mathbf{x}(t) ; \Theta)-\mu_{R}(t)}{\sigma_{R}(t)},
\end{equation}in this formulation, \(\mathbf{x}(t)\) represents the aggregated features of the edge at time \(t\), \(\mathbf\Theta\) denotes the weights and biases of the multilayer perceptron, \(\mathbf\phi(\mathbf{x}(t) ; \Theta)\) represents the anomaly score of the event , while \(\mu_{R}(t)\) and \(\sigma_{R}(t)\) denote the mean and standard deviation of the reference scores at that time, respectively.

\begin{equation}
\begin{aligned}
L_{dev}\left(\phi(\mathbf{x}(t) ; \Theta), \mu_{R}(t), \sigma_{R}(t)\right) &= 
(1 - y) \cdot (\operatorname{dev}(\mathbf{x}, t))^{2} \\
&\hspace{-2.5em} + y \cdot \operatorname{sigmoid}(a - \operatorname{dev}(\mathbf{x}, t)).
\end{aligned}
\end{equation}

Here, \(y\) serves as the predictive label for the edge at time \(t\), and \(\alpha\) approximates the time-dependent confidence interval. We modified the deviation term into a squared form to enhance the reciprocal effect of the loss function. Compared to the max function, the sigmoid function provides a smoother gradient within a certain range, which helps stabilize the distance between the overall distribution of the learning memory library and the anomaly scores of normal edges.

To integrate edge reconstruction with abnormal structure detection during the model training process, TFLAG combines \(L_{\text {pred }}\) and \(L_{\text {dev}}\) to construct a comprehensive loss function. This function ensures that as the model predicts edge categories, it also learns the normal and abnormal reconstruction information of edges. The total loss function is defined as follows: \begin{equation}
\label{ex8}
L_{\text {all }}=\lambda_{1} L_{\text {pred }}+\lambda_{2} L_{d e v},
\end{equation}
the constants \(\lambda_{1}\) and \(\lambda_{2}\) are hyperparameters used to balance the contributions of the two loss functions.

\subsection{Anomaly Detection}

The TFLAG system adopts a time window-level granularity for anomaly detection. Events occurring within a fixed-size event window \(T\) at time \(t\) comprise 1) the attributes and categories of the source and destination nodes, 2) the reconstruction error of the edge, and 3) the deviation loss of the anomaly score. TFLAG segments the data under investigation into equidistant time windows, conducts anomaly detection based on the attributes of the model output within each window, and identifies the window with the highest anomaly value as the central anomaly window. Surrounding this central window, abnormal edge windows are extracted and linked over multiple periods to construct a complete attack incident.

\textbf{Anomaly Window Detection. }Previous studies employing node frequency-based methods for measuring rarity were overly simplistic, generally focusing solely on the frequency of node occurrences in the data or the path's branch count \cite{ref31,ref32}. This approach overlooks fundamental anomaly analysis and the influence of temporal structure. Although the state-of-the-art method \cite{ref10} takes both factors into consideration, its practical application faces challenges: when connecting multiple anomaly windows, it is necessary to precompute the node frequency of event attributes and rely on prior knowledge to manually exclude event attributes that may influence the final calculation of anomaly scores, which is difficult to achieve in real-world scenarios. In contrast, the anomaly detection module of TFLAG is designed based on the low frequency and continuity of APT attacks, so it does not require the prior exclusion of normal event attributes that may influence the calculation, and can directly detect the complete attack sequence. 

The operational steps include:
\textbf{Initial Anomaly Ranking. }Time windows are prioritized based on the events that exhibit the highest anomaly reconstruction loss, resulting in a sequence of windows ranked in descending order of reconstruction loss.
\textbf{Refining Anomaly Detection. }Following initial sorting, the top 10\% of windows undergo further refinement using the deviation loss of anomaly scores. An edge characterized by high reconstruction loss yet average deviation loss is likely indicative of normal behavior that diverges from typical system activity patterns.
\textbf{Frequency-Based Filtering. }An integral part of attack filtering involves analyzing event frequency to pinpoint the central anomalous window. APT attacks are typically marked by elusive, low-frequency behaviors, contrasting with high node frequencies, which usually signify the routine operations of specific system processes.

\textbf{Anomaly Events Detection. }To effectively detect the multi-stage, continuous behaviors characteristic of APT attacks, it is crucial to utilize the distinguishing features of anomalous events and nodes identified in the previously determined central anomalous window. We then extend the search sequentially, first forward and then backward in time, to delineate the entire timeframe of the attack.
\begin{equation}
\label{ex9}
S_{\mathcal{T}_{i}} \cap S_{\mathcal{T}_{i+1}} \neq \emptyset \quad OR \quad E_{\mathcal{T}_{i}} \cap E_{\mathcal{T}_{i+1}} \neq \emptyset.
\end{equation}


The terms \(S_{T_{i}}\) and \(E_{T_{i}}\) represent the set of nodes and the set of edges within the time window. When the specified conditions are satisfied, the two windows are deemed relevant to the attack event. Subsequently, we rank the edges based on reconstruction losses and employ Jaccard similarity \cite{ref33} to assess node similarity. This metric is then compared against the last two windows, and if the results remain inconclusive, a feature value is established as the baseline for anomaly detection.

\begin{equation}
\label{ex10}
Anomaly Jaccard =\frac{\left|SX_{T_{i}} \cap SX_{T_{i+1} \mid}\right|}{\left|S_{T_{i}} \cup S_{T_{i+1}}\right|}.
\end{equation}

The terms \(SX_{T_{i}}\) and \(SX_{T_{i+1}}\) refer to sets of low-frequency nodes extracted from two adjacent time windows. The Jaccard similarity, which ranges from 0 to 1, quantifies the degree of similarity between these sets; a value closer to 1 signifies a stronger similarity.

By calculating the Jaccard similarity among anomalous low-frequency nodes, we can assess their overlap, enabling a more precise identification of anomalous events. A higher score implies a stronger correlation, suggesting that the nodes belong to the same attack. Therefore, the TFLAG anomaly detection system obviates the necessity for prior knowledge of noise nodes, capitalizing on the stealthy characteristics of APT attacks to enhance detection accuracy and reduce false positives.

\subsection{Model Adaption}
APT anomaly detection presents considerable challenges due to the covert and complex nature of such threats, which typically lack clear patterns. The pervasive problem of concept drift in anomaly detection highlights the necessity of assessing its impact on actual systems. The TFLAG system, a self-supervised learning framework, leverages edge prediction to reconstruct the model without the need for labeled data, thereby enhancing anomaly detection capabilities. Despite its innovation and efficacy, this method may produce a certain number of false positives in real-world scenarios when benign behaviors, which deviate from normal system patterns, are erroneously classified as anomalies. To mitigate this issue, the TFLAG system employs a multi-category screening mechanism that effectively lowers the false alarm rate by distinguishing between true anomalies and normal variations in behavior.

In anomaly detection, the principal challenge posed by concept drift is accurately distinguishing whether abrupt deviations in normal behavior represent false positives \cite{ref49,ref50}. Although these behaviors may appear anomalous, they do not necessarily indicate a malicious attack. To address this, the TFLAG system implements two strategies: dynamically adjusting anomaly score distributions through an abnormal score memory storage module and incorporating new behavioral feedback into the model. These continuous adjustments ensure the system consistently adapts to emerging patterns while preserving detection accuracy. Furthermore, the system regularly updates the data within the memory bank to maintain its relevance and effectiveness. As a result, these modifications enhance the model's capacity to differentiate between novel and routine data, thereby bolstering its resilience against APT in the context of concept drift.

\section{EXPERIMENTS AND ANALYSIS}
Our system was developed using Python. Log parsers were constructed for various datasets, and a unified format for graph access was implemented in the database. The graph learning framework utilizes a multi-layer TGAT \cite{ref15} and PyTorch \cite{ref34}. Evaluations were conducted across four public datasets, employing two detection strategies: graph-level and window-level. All experiments were performed on a server running Ubuntu 22.04.4 LTS with kernel version 5.15.0-117-generic, equipped with an Intel Xeon Platinum 8352V processor (16 cores at 2.10 GHz) and 64 GB of RAM. Within the TFLAG system settings, the graph encoder's output dimension is set at 128, the training batch size at 512, the number of neighbor nodes at 20, and the learning rate at 0.0001. Subsequent experiments examined the influence of varying hyperparameters on performance, with selections including neighbor size \textit{N} from {5, 10, 20, 30, 40}, hidden layer dimension \textit{h} from {32, 64, 128, 256}, edge embedding dimension \textit{e} from {4, 8, 16, 32}, and time window size \textit{T} from {5, 10, 15, 20}. In Section V-A, we introduce the datasets utilized by the TFLAG system at both graph and window levels. Section V-B details the system's effectiveness across different scenarios, comparing it to the latest research on the same granularity. Section V-C provides additional comparisons, while Section V-D offers an in-depth comparison of the impact of various TFLAG hyperparameters on the system's detection outcomes. The entire set of experiments robustly demonstrates the TFLAG system's precision in detecting anomalies under unsupervised conditions.

\subsection{Datasets}

We evaluate the effectiveness of TFLAG on multiple public datasets: StreamSpot \cite{ref17} and DARPA \cite{ref16,ref51}. These datasets vary in granularity, logging formats, attack methods, and sources. Verifying the effectiveness of the system on these diverse datasets can reflect the performance of TFLAG, and currently the most advanced APT detection methods are also tested on these datasets, allowing for effective result comparisons. The composition and source format of the datasets are introduced in detail below.

\begin{table} 
\setlength{\abovecaptionskip}{-10pt}
\renewcommand\arraystretch{1.1}
	\caption{Overvier of StreamSpot Dataset}        
	\label{table0}           
	\begin{center}                 
		\begin{tabular}{m{1.8cm}  m{1.8cm}<{\centering} m{1.8cm}<{\centering} m{1.8cm}<{\centering}}
			\toprule               
			Scenario&ALL \#Nodes&ALL \#Edges&Size(GB)\\       
			\hline                 
			Dwnload&8,830&310,814&1.0 \\
			Gmail&6,826&37,382&0.1\\
                VGame&8,636&112,958&0.4 \\
                YouTube&8,292&113,229&0.3\\
                Attack&8,890&28,423&0.1\\
                CNN&8,989&294,903&0.9\\
			\bottomrule            
		\end{tabular}
	\end{center}
\vspace{-2.0em}
\end{table}

\textbf{StreamSpot Datasets.}
This dataset provides provenance graph data by simulating a laboratory environment, capturing normal and attack behaviors across six scenario graphs. Each scenario graph comprises 100 graphs that record the underlying process traffic monitored in a time series. As detailed in Table \ref{table0}, five of these scenarios (YouTube, Gmail, Video Game, Attack, Download, CNN) represent normal user behaviors and are entirely benign; the remaining scenario consists of 100 graphs depicting abnormal behavior generated by simulating driver download attacks. It is important to note that this dataset focuses on graph-level anomaly detection rather than fine-grained window or entity-level detection. Following its previous methodologies, the TFLAG method performs graph-level detection on this dataset. Each scenario graph is composed of chronologically arranged edges, aligning with the TFLAG premise of training on benign datasets and testing on abnormal ones. For our approach, one graph from each benign scenario is used for training, while the rest of the benign graphs and all abnormal graphs are utilized for testing and evaluation. This method enables us to determine whether each graph is anomalous and to benchmark our detection results against current state-of-the-art APT detection methods.

\begin{table} 
\setlength{\abovecaptionskip}{-10pt}
\renewcommand\arraystretch{1.1}
	\caption{Overvier of DARPA Dataset}        
	\label{table1}           
	\begin{center}                 
		\begin{tabular}{m{1.8cm} m{2.2cm}<{\centering} m{1cm}<{\centering} m{1cm}<{\centering} m{0.8cm}<{\centering}}
			\toprule               
			Dataset &Anomaly Scenario&ALL \#Nodes&ALL \#Edges&Size(GB)\\       
			\hline                 
			CADETS-E3 &Nginx Backdoor&613,713&29,727,532&17.9 \\
			THEIA-E3 &Browser Extension&1,259,368&44,066,112&81.9\\
            ClearScope-E3 &Firefox Backdoor&201,021&8,915,438&16.6 \\
            CADETS-E5 &Nginx Drakon&394,617&77,886,133&106.9\\
            THEIA-E5 &Firefox Drakon&1,511,241&74,565,731&140.3\\
            ClearScope-E5 &Lockwatch APK&268,251&34,413,590&67.7\\
			\bottomrule            
		\end{tabular}
	\end{center}
\vspace{-2.0em}
\end{table}

\textbf{DARPA Datasets.}
The DARPA dataset originates from the third and fifth Transparent Computing (TC) adversarial exercises, as shown in Table \ref{table1}. It encompasses dataset generation techniques, scenario descriptions, tools used, attack details, evaluations of each analysis report, sample attack graphs, and logs. The attacker's goal was to steal proprietary and personal information from a target company by exploiting a \texttt{FreeBSD} web server, injecting the \texttt{SSHD} process, and monitoring network activity to exploit discovered hosts and extract data. Ultimately, the attacker targeted a mobile device to extract personally identifiable information (PII), using tools such as Nginx and Firefox backdoors, browser extensions, and Drakon APT. Additionally, the attack involved HTTP communication, creating elevated processes, and injecting \texttt{.dll} or \texttt{.so} files. Multiple logging systems captured host activity, documenting underlying process traffic throughout the exercise.

\begin{table} 
\setlength{\abovecaptionskip}{-10pt}
\renewcommand\arraystretch{1.1}
    \caption{Dataset Label Division}        
    \label{table2}           
    \begin{center}                 
        \begin{tabular}{m{1.8cm} m{2.1cm}<{\centering} m{1.8cm}<{\centering} m{1.7cm}<{\centering}} 
            \toprule               
            Dataset & Training Data (yyyy-mm-dd) & Validation Data (yyyy-mm-dd) & Test Data (yyyy-mm-dd) \\ 
            \hline                 
            CADETS-E3 & 2018-04-03/04/05 & 2018-04-07 & 2018-04-06 \\
            THEIA-E3  & 2018-04-03/04/05 & 2018-04-09 & 2018-04-10 \\
            ClearScope-E3 & 2018-04-07/08/09 & 2018-04-10 & 2018-04-11 \\
            CADETS-E5 & 2019-05-08/09/11 & 2019-05-17 & 2019-05-16 \\
            THEIA-E5  & 2019-05-08/09/11 & 2019-05-14 & 2019-05-15 \\
            ClearScope-E5 & 2019-05-08/09/11 & 2019-05-15 & 2019-05-17 \\
            \bottomrule            
        \end{tabular}
    \end{center}
\vspace{-2.0em}
\end{table}

We evaluated the TFLAG method using the underlying log data obtained from different capture systems \cite{ref16} in the adversarial exercises, extracting critical edges and nodes essential for model training and anomaly detection. As shown in Table \ref{table2}, to assess the model's entity-level detection capabilities on this large, unlabeled dataset effectively, we implemented a daily training regimen. Specifically, with the CADETS-E3 dataset, we preprocessed and trained using the first two days of normal data. We used the third day’s benign data for evaluation and the fourth day’s data, which included abnormal attacks, for testing. This approach allowed us to evaluate authentic attack behaviors and benchmark detected abnormal event windows against the DARPA dataset's ground truth, enabling us to accurately identify the critical exceptional window and correlate all related anomalous events without prior knowledge.

\begin{figure}
\centering
\includegraphics[width=3.5in]{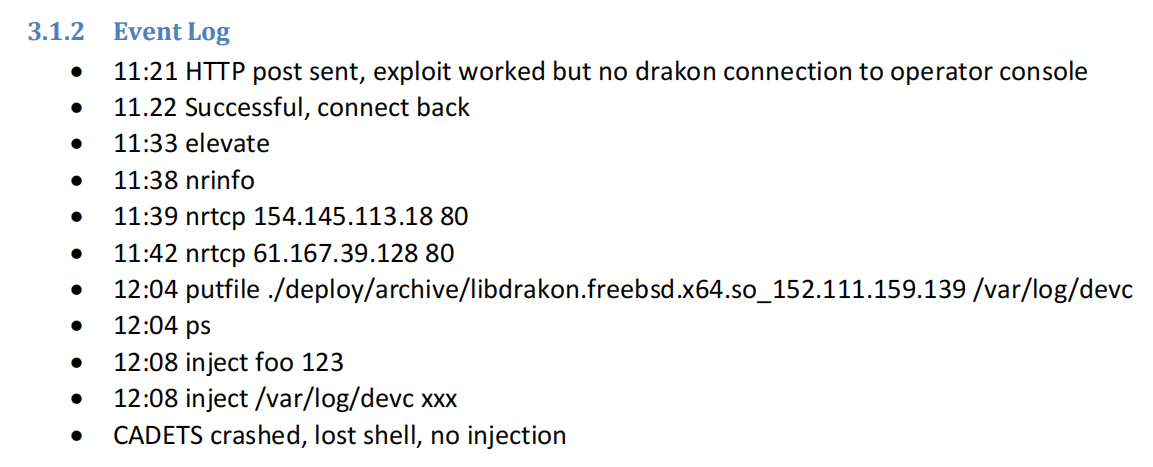}
\caption{Specific attack time periods are provided in the Ground\_Truth\_Report file of the official DARPA dataset.}
\label{fig_4}
\vspace{-1.0em}
\end{figure}

As highlighted in Fig.~\ref{fig_4}, the DARPA dataset provides detailed information on attack events, including specific timestamps and behavioral descriptions. This data facilitates the precise alignment of detected anomalous event windows with actual attack periods. Unlike previous methods that relied on manual annotation \cite{ref35}, our approach does not require predefined label information and can independently identify the core abnormal activity window solely by learning from benign data. Additionally, by performing anomaly detection on the dataset, we can compare our results with the current SOTA systems in anomaly-based provenance graph detection, thus validating the effectiveness of the TFLAG method.

\subsection{Effectiveness}

We evaluated the effectiveness of the TFLAG method for detecting APT attacks across multiple scenario datasets at varying levels of granularity. To emulate real-world scenarios, training was conducted on days without anomalies in chronological order for each dataset, using the days with anomalies and their adjacent days as the subjects of detection. We conducted experiments on the SteamSpot dataset and DARPA, and the detailed results of these analyses are presented in Table \ref{table3}, where the evaluation metrics used are consistent with those commonly employed by other methods in the field.

Precision, recall, accuracy, and the area under the receiver operating characteristic curve (AUC) for the datasets were computed based on predefined time windows. The attack log files provided were manually annotated to differentiate between bengin and attack-related time windows. If a time window was inaccurately labeled as anomalous by the TFLAG method, it was classified as a false positive (FP); conversely, correct identification of an attack window by the TFLAG method was considered a true positive (TP). The instances of false negatives (FN) and true negatives (TN) were calculated using the same rationale. Table \ref{table3} also enumerates the specific counts of TP, TN, FP, and FN for each dataset. The results indicate that the TFLAG method successfully identified all abnormal windows in the DARPA-E3 and StreamSpot datasets, but exhibited false negatives in the ClearScope-E3 and THEIA-E3 datasets, and false positives in the DARPA-E5 dataset. The details of these are further dissected below.

Upon further examination of the missed time windows, we identified specific limitations in the TFLAG method's detection capabilities within the THEIA-E3 and ClearScope-E3 datasets. In THEIA-E3, TFLAG erroneously labeled the initial time window of the attack sequence as benign due to a system lock-up, which caused the system to become unresponsive and delayed detection. During this period, the attacker attempted to deploy attack code at the \texttt{www.allstate.com} website to deliver a malicious payload, but the system continued to emit stale data, leading TFLAG to incorrectly classify the timeframe as benign. Once the system regained stability and resumed normal operations, subsequent attack efforts were correctly logged. Similarly, in ClearScope-E3, TFLAG failed to recognize the final time window of the attack, during which the attacker made unsuccessful attempts at privileged file operations. Although the system returned to normal after the attack, the dominance of normal activity in this window led TFLAG to mistakenly classify it as non-anomalous. These detection errors by TFLAG can generally be attributed to its sensitivity to the timing and extent of manifest attack behaviors within the designated windows. However, it is important to note that these issues do not impact the administrator's ability to detect the attack or the overall APT attack analysis, as the attack sequence was eventually detected, and the overall analysis remains intact.

\begin{table} 
\setlength{\abovecaptionskip}{-5pt}
\renewcommand\arraystretch{1.1}
    \caption{Detection Experiment Results of Selected Datasets} 
    \label{table3} 
    \begin{center} 
        \begin{tabular}{m{1.75cm} m{0.25cm}<{\centering} m{0.25cm}<{\centering} m{0.2cm}<{\centering} m{0.2cm}<{\centering} m{0.7cm}<{\centering} m{0.5cm}<{\centering} m{0.7cm}<{\centering} m{0.5cm}<{\centering}} 
            \toprule 
            Datasets & TP & TN & FP & FN & Precision & Recall & Accuracy & AUC \\ 
            \hline 
            StreamSpot & 100 & 375 & 0 & 0 & 1.000 & 1.000 & 1.000 & 1.000 \\
            CADETS-E3 & 4 & 181 & 0 & 0 & 1.000 & 1.000 & 1.000 & 1.000 \\
            ClearScope-E3 & 4 & 180 & 0 & 1 & 1.000 & 0.800 & 0.995 & 0.900 \\
            THEIA-E3 & 7 & 117 & 0 & 1 & 1.000 & 0.875 & 0.992 & 0.938 \\
            
            CADETS-E5 & 6 & 246 & 3 & 1 & 0.667 & 0.857 & 0.984 & 0.923 \\
            ClearScope-E5 & 10 & 240 & 4 & 1 & 0.714 & 0.909 & 0.800 & 0.946 \\
            THEIA-E5 & 2 & 180 & 1 & 0 & 0.667 & 1.000 & 0.994 & 0.997 \\
            \bottomrule 
        \end{tabular}
    \end{center}
\vspace{-2.0em}
\end{table}

\begin{table*}[] 
\setlength{\abovecaptionskip}{-5pt}
\renewcommand\arraystretch{1.2}
    \caption{TFLAG and KAIROS examine the distinctive attributes of the most anomalous edges within the same anomaly window and compare them with actual attack scenarios.} 
    \label{table4} 
    \begin{center} 
        \begin{tabular}{m{1cm} m{2.7cm} m{1cm}<{\centering} m{3cm} m{4.7cm} m{1.5cm}<{\centering}} 
            \toprule 
            Datasets & Audit Log Anomaly & System & Anomaly Time Window & Anomaly Edge & Link Loss \\ 
            \hline 
            CADETS   & F1-elevate /tmp/vUgefa  & KAIROS & 4-6-11:33$\sim$4-6-11:48 & netflow:128.55.12.10:53 → subject: lsof & 7.4449 \\
            CADETS   & F1-elevate /tmp/vUgefa & TFLAG    & 4-6-11:30$\sim$4-6-11:45 & subject:vUgefal → file:/tmp/vUgefal & 20.7332 \\
            
            \bottomrule 
        \end{tabular}
    \end{center}
\vspace{-2.0em}
\end{table*}

In the DARPA-E5 scenario, the attacker initially exploited a vulnerable process before proceeding to target additional processes. Consequently, previously attacked components were mistakenly identified as attack events upon resuming normal system behavior. As an anomaly-based detection system, TFLAG demonstrates improved sensitivity to both false positives and genuine attacks that deviate from established norms. This resulted in a misclassification, wherein previously compromised events were erroneously categorized as normal behavior. To effectively mitigate false positives associated with normal system behavior, it is essential to implement re-labeling and retraining in subsequent stages.

\begin{figure}[thbp!]
    \centering
    \subfloat[Kairos reconstruction loss distribution]{\includegraphics[width=3in]{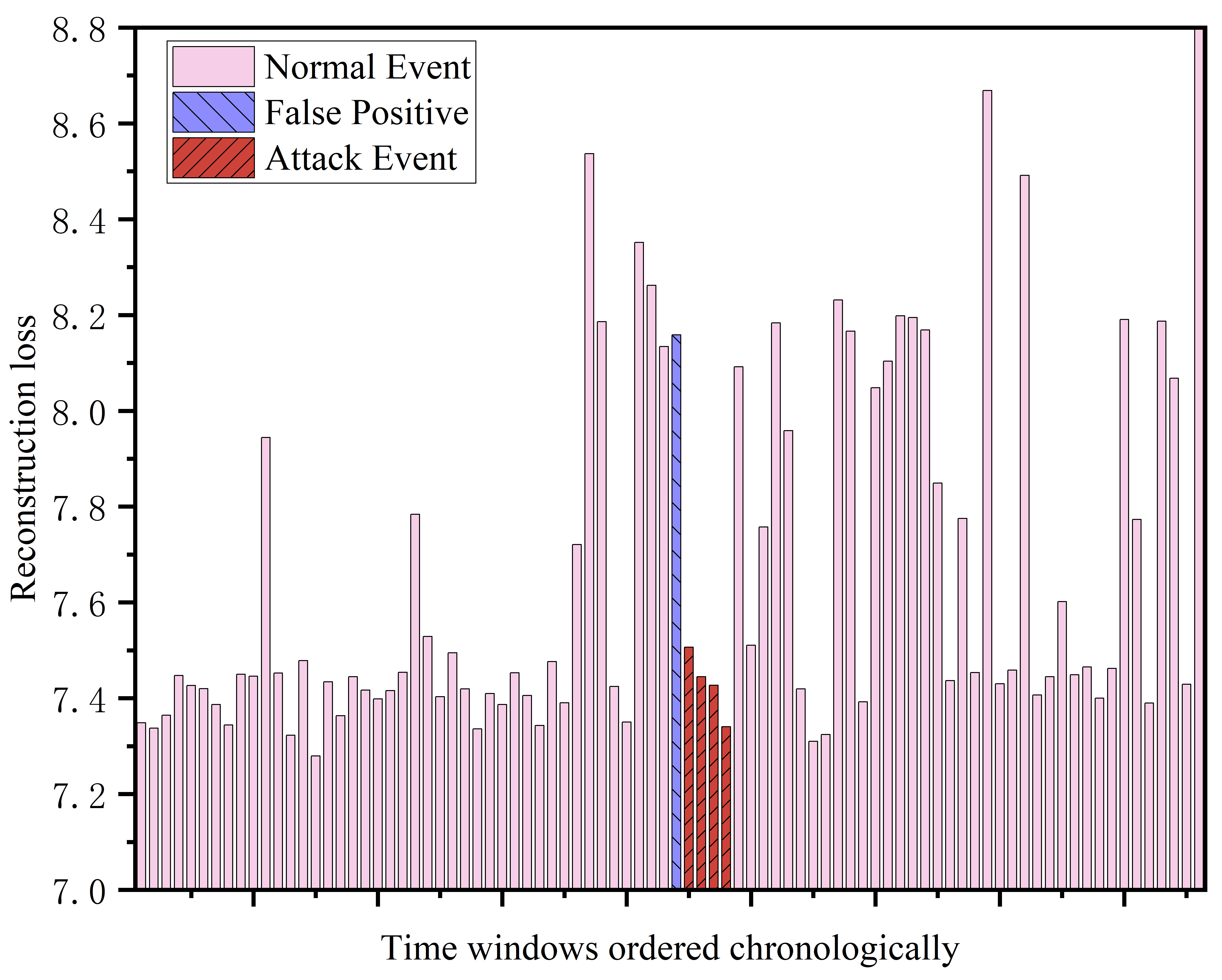}\label{kairos_loss}}\\
    \subfloat[TFLAG reconstruction loss distribution]{\includegraphics[width=3in]{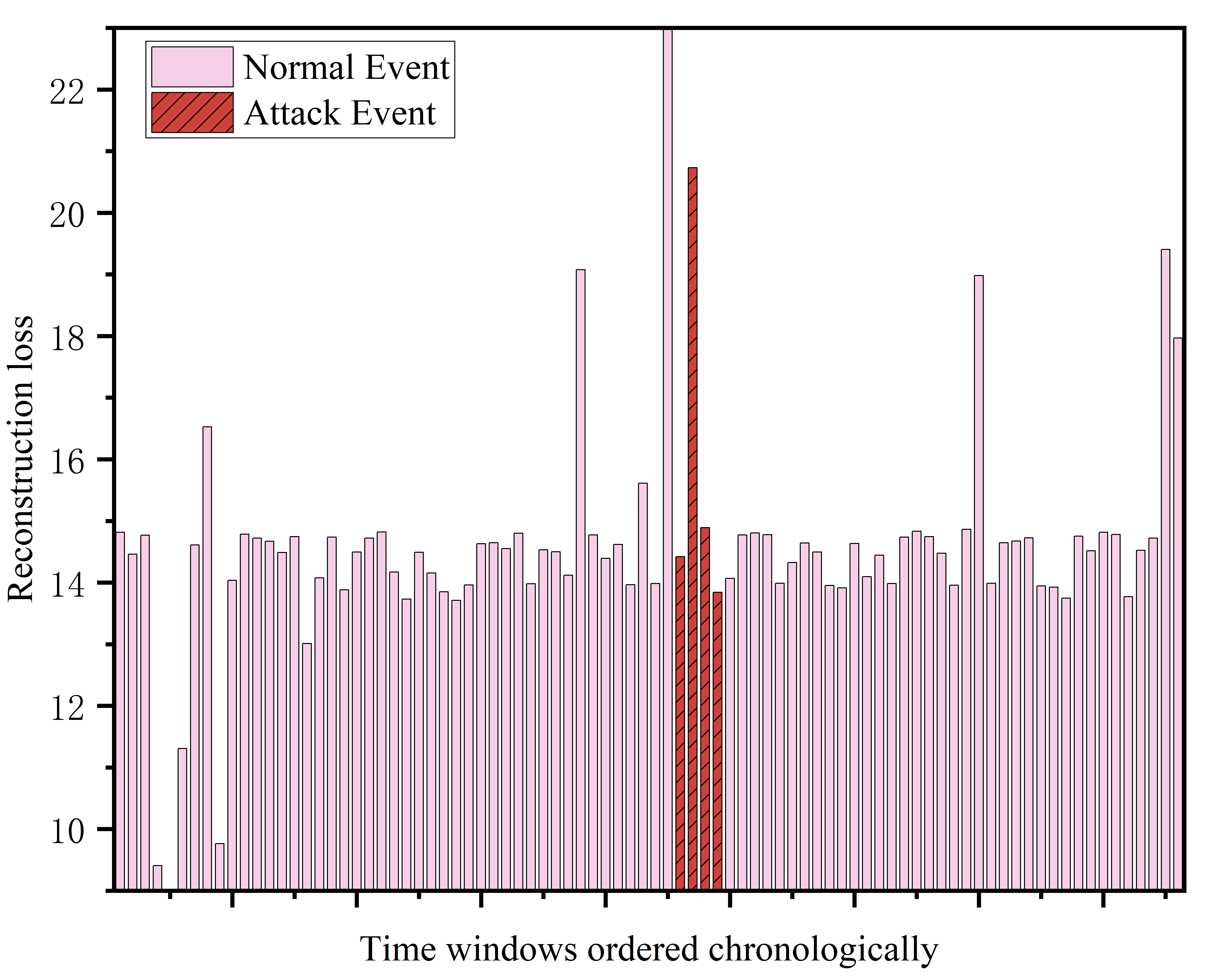}\label{TFLAG_loss}}\\
    \caption{The distribution of the highest reconstruction loss events for each time window in the detection results of 2018-04-06, for both Kairos and TFLAG in CADETS-E3 is presented.}
\vspace{-1.0em}
\label{loss}
\end{figure}

\begin{figure}
\centering
\includegraphics[width=3in]{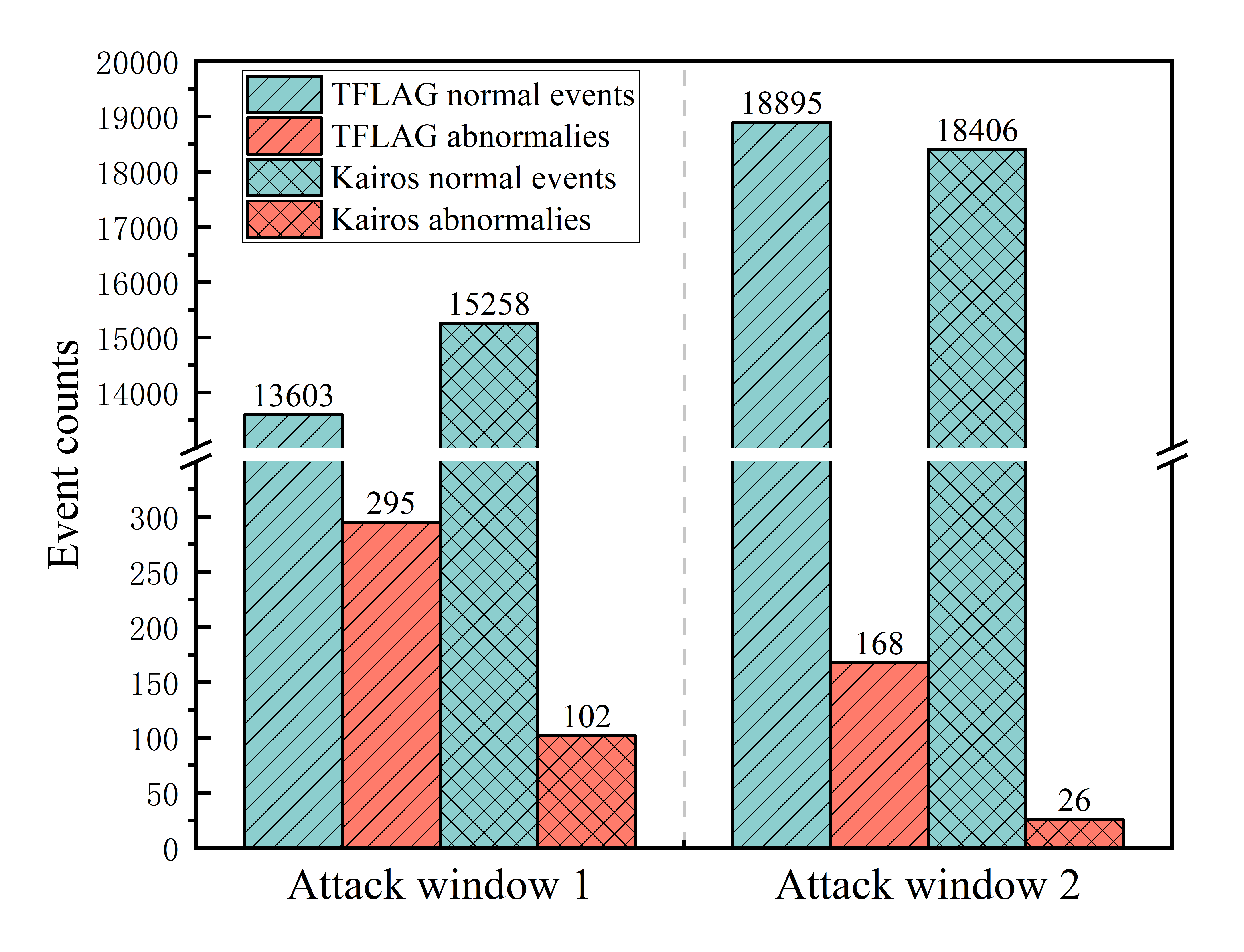}
\caption{Comparison of true anomalous events among the top 10\% reconstruction loss events for TFLAG and KAIROS within the same anomalous time window.}
\label{events}
\vspace{-1.0em}
\end{figure}

\subsection{TFLAG vs. SOTA.}

The datasets used to evaluate the TFLAG method are similarly applied to assess SOTA APT detection methods. Both TFLAG and KAIROS employ a label-free, self-supervised anomaly detection system, as opposed to the label-based PIDS\cite{ref6,ref36,ref37,ref38}. Therefore, direct comparisons with more advanced methodologies such as FLASH, MAGIC, HOLMES, and JBEIL, which depend on project code analysis and abnormal node labels from TreaTrace or manually generated labels, are impractical. While numerous studies examine anomaly-based detection of APT, the detection granularity and labeling used for comparison in their systems differ from those in our approach. According to its publications\cite{ref10}, \textbf{KAIROS has surpassed these methods in terms of detection efficiency}. Consequently, this study concentrates on comparing TFLAG with KAIROS to ascertain the effectiveness of our method in detecting APT attacks. This comparison clarifies TFLAG’s performance and advantages in unsupervised APT detection, independent of labeled dataset quality control concerns.

KAIROS utilizes an encoder-decoder framework within a Temporal Graph Neural Network (TGN) to detect anomalies by reconstructing network edges, encoding neighborhood structures, and updating them via a memory module. Reproduction of the code revealed that, during its detection phase, numerous false positives associated with system behavior result in a higher recall rate but a reduced accuracy rate. 
In contrast, TFLAG demonstrates higher precision in detecting anomalous edge events. As shown in Fig.~\ref{loss}, within the CADETS-E3 scenario, TFLAG successfully detected the most significant anomalous loss during the attack event window. Specifically, within the first fifteen minutes after the attack began, KAIROS generated a false positive with notable anomalous activity, which was not an actual attack. In the following four time windows, the detected loss values did not show a significant increase. KAIROS later identified the attack queue after filtering out noise nodes and analyzing node rarity, but this was done with the false positive window still present. In contrast, TFLAG accurately captured the high-anomaly events within the attack window, where the anomaly score was significantly higher than the maximum values in other time windows.
Additionally, as observed in the figure, TFLAG also successfully identified high-loss events in the previous time window. By analyzing the deviation scores, the rarity of the anomalous events, and the loss values, it effectively filtered out this false-positive time window. This precise anomaly detection method allows TFLAG to provide more reliable attack identification. Table \ref{table4} shows that the anomalous events detected by TFLAG align highly with the actual records in the E3 attack dataset.

Under the same training conditions and time windows, TFLAG accurately identified the malicious entity \textbf{vUgefal} writing the file \textbf{/tmp/vUgefal} within the attack window, recording a peak reconstruction loss of \textbf{20.7}. This peak clearly distinguished it from other events within the same time window. In contrast, KAIROS erroneously classified a benign system behavior, specifically the \textbf{lsof} process checking the network \textbf{128.55.12.10:53}, as the most anomalous event. This event was not mentioned in the attack report, but its reconstruction loss was the highest in the entire time window, reaching \textbf{7.4}. This impacted subsequent administrator trace-back decisions.
Furthermore, the construction of the deviation network effectively differentiates between attack events and false alarm events, as illustrated in Fig.~\ref{events}. Within the same attack window, we can identify a greater number of genuine attack events among the top 10\% of anomalous events. This capability allows administrators to more accurately pinpoint rare attack events exhibiting a higher degree of anomaly during subsequent fine-grained window-based trace analysis.

Table \ref{table5} demonstrates that TFLAG consistently surpasses KAIROS in accuracy across all three window detection levels. Moreover, in the node detection level dataset SteamSpot, both TFLAG and KAIROS achieve an exceptional 100\% accuracy and recall. In the E3 dataset, TFLAG correctly identified the entire attack event within the CADETS anomaly detection window, achieving 100\% accuracy. Conversely, KAIROS erroneously categorized a prior window as anomalous, detracting from its accuracy. Similar observations were made in the ClearScope and THEIA datasets; although our method occasionally misses anomaly windows at the beginning or end of an attack, it does not erroneously identify benign windows as anomalous, primarily due to its enhanced sensitivity to anomaly edges compared to KAIROS. Overall, TFLAG’s methodology for identifying anomaly windows far surpasses that of KAIROS, reducing false positives and improving both recall and accuracy rates, thus proving more effective and robust in APT detection.

\begin{figure*}[thbp!]
    \centering
    \subfloat[\(N\)]{\includegraphics[width=1.6in]{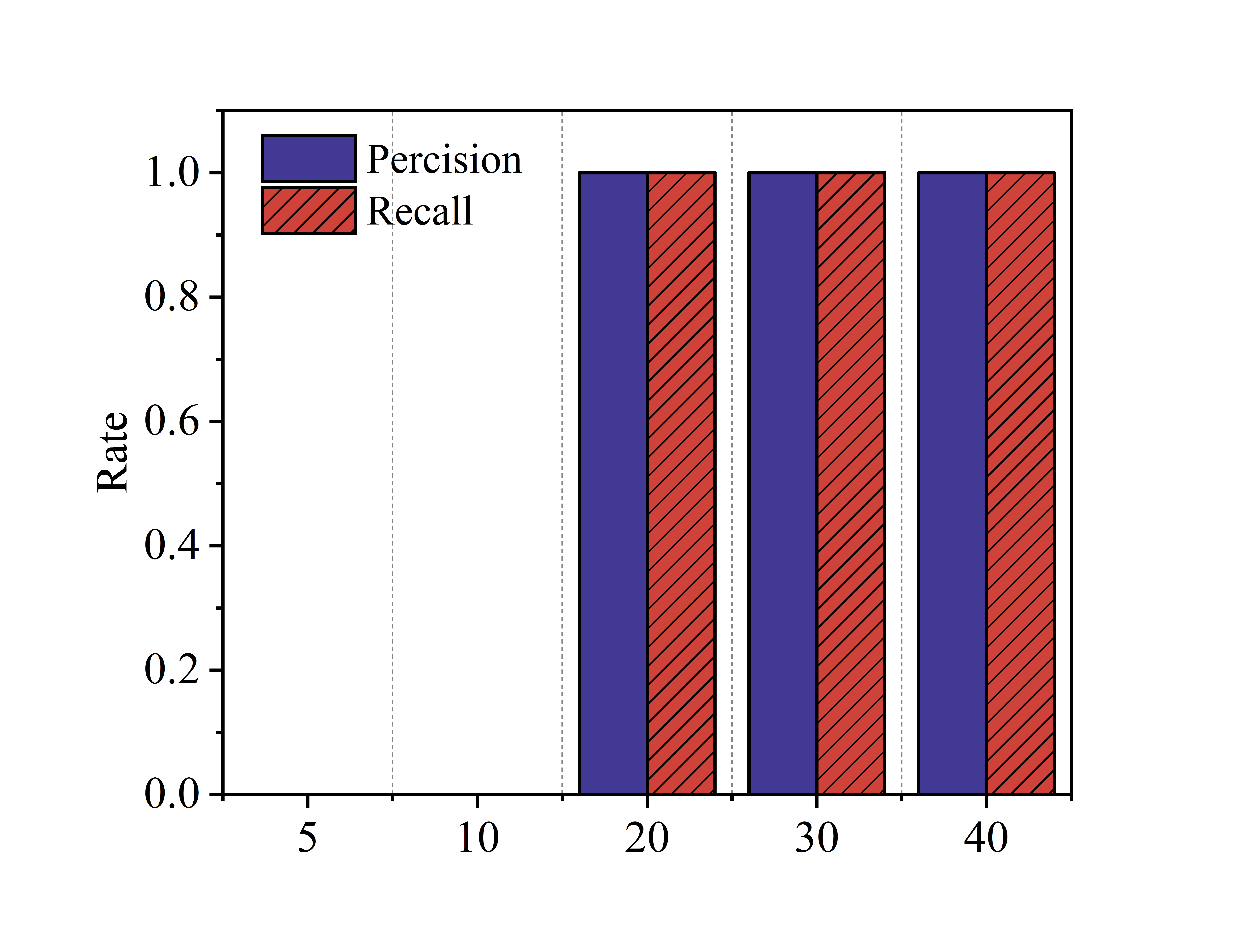}\label{fig:N}}
    \hfil
    \subfloat[\(h\)]{\includegraphics[width=1.6in]{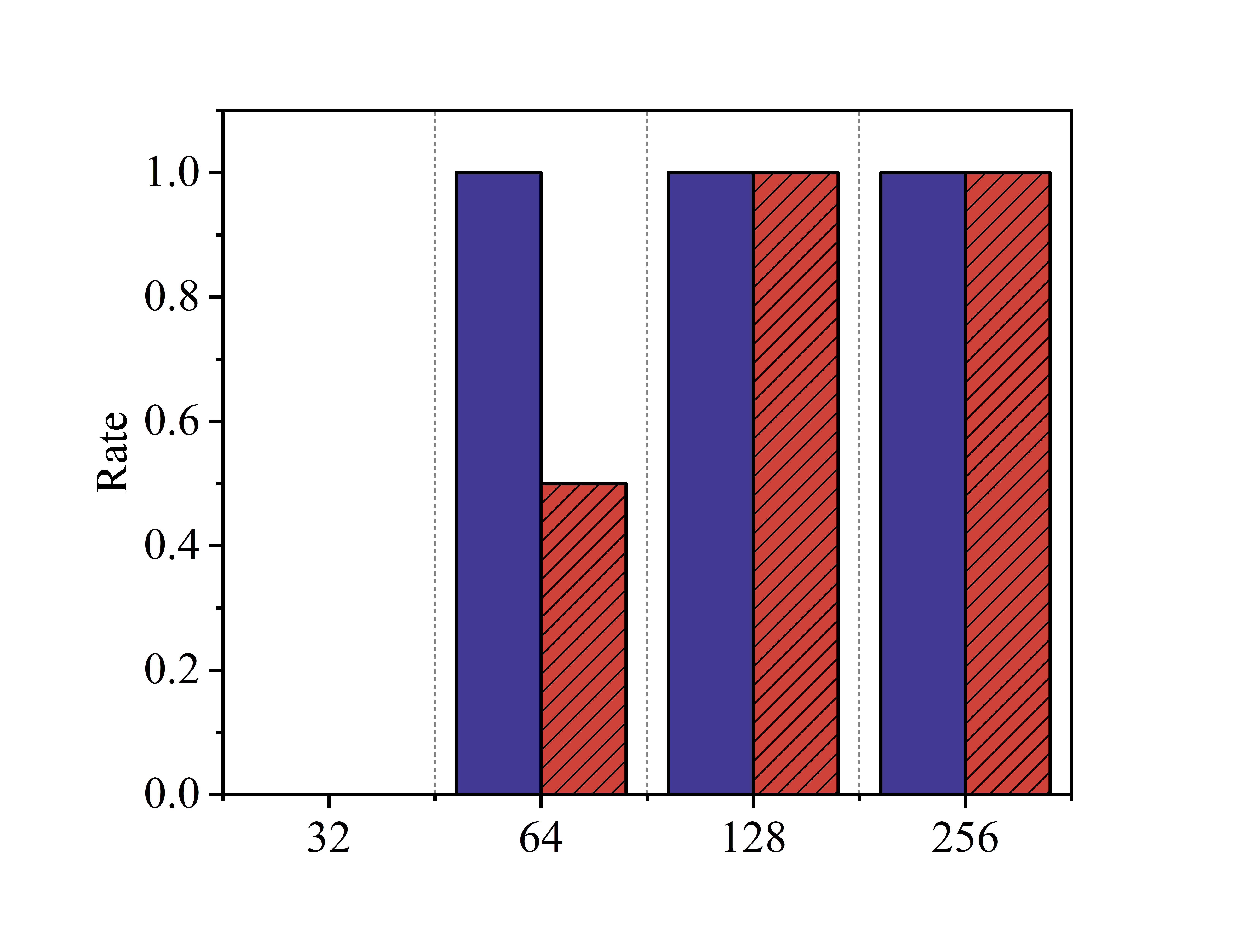}\label{fig:h}}
    \hfil
    \subfloat[\(e\)]{\includegraphics[width=1.6in]{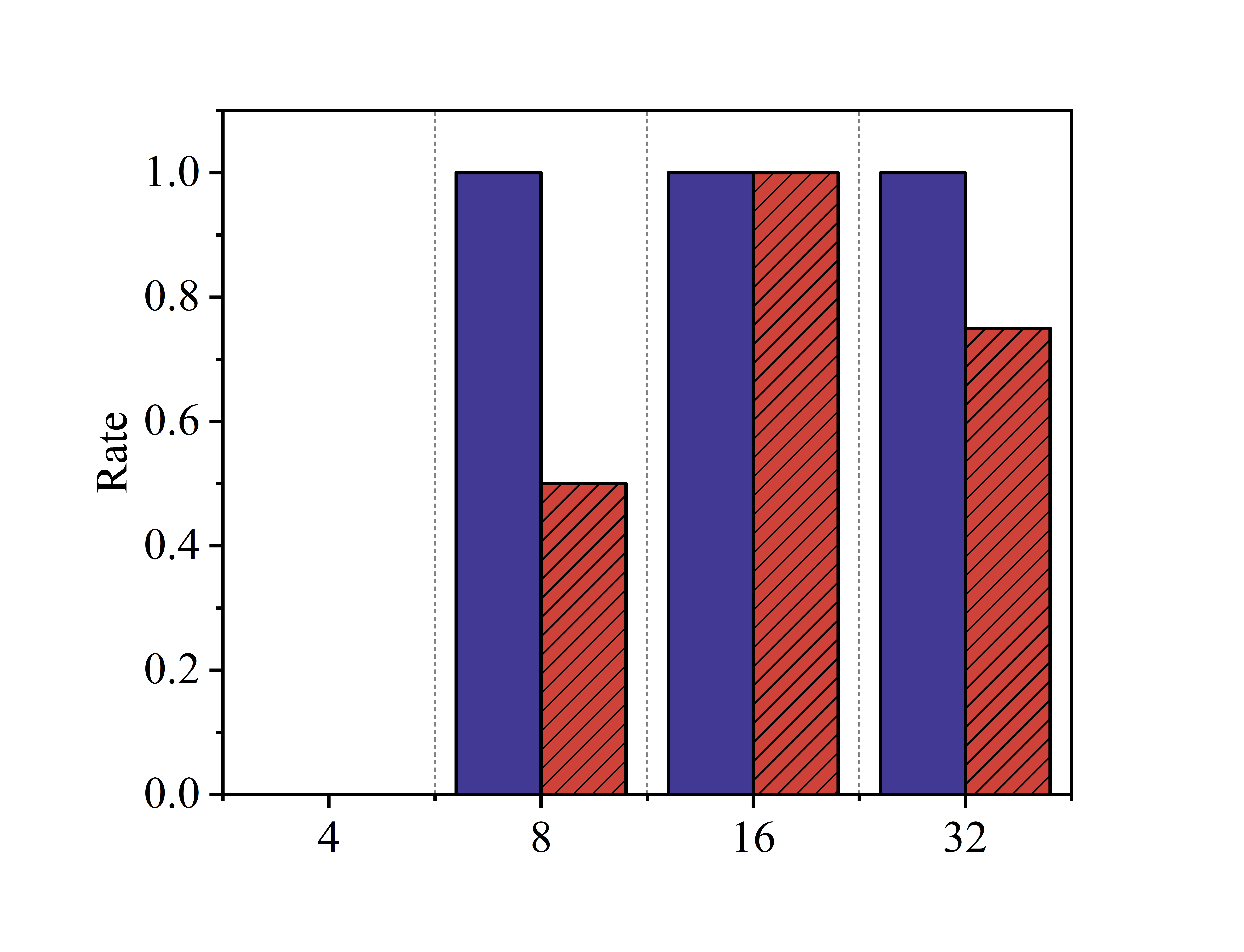}\label{fig:e}}
    \hfil
    \subfloat[\(T\)]{\includegraphics[width=1.6in]{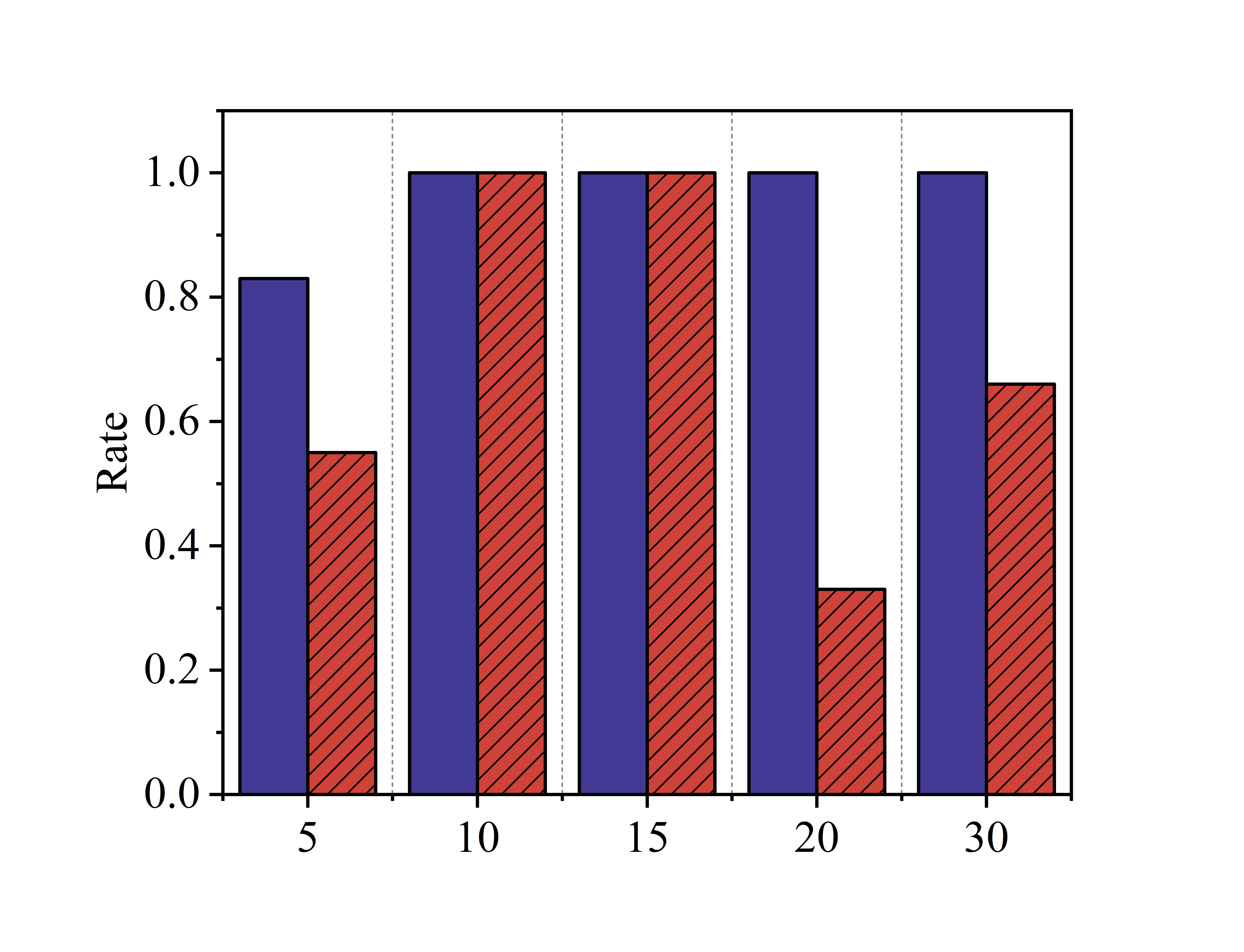}\label{fig:T}}
    \caption{The detection performance of the TFLAG method within the E3-CADETS attack scenario is assessed by analyzing the impact of various hyperparameter values on accuracy and recall, while holding other parameters constant.}
\vspace{-1.0em}
\end{figure*}

\begin{table} 
\setlength{\abovecaptionskip}{-5pt}
\renewcommand\arraystretch{1.2}
    \caption{Comparison study between TFLAG and KAIROS} 
    \label{table5} 
    \begin{center} 
        \begin{tabular}{m{1.5cm} m{1.5cm}<{\centering} m{1cm}<{\centering} m{1cm}<{\centering} m{1.2cm}<{\centering}} 
            \toprule 
            Datasets & System & Precision & Recall & Accuracy \\ 
            \hline 
            StreamSpot & KAIROS & 1.000 & 1.000 & 1.000 \\
                        & TFLAG    & 1.000 & 1.000 & 1.000 \\
            CADETS     & KAIROS & 0.800 & 1.000 & 0.994 \\
                        & TFLAG    & \textbf{1.000} & 1.000 & \textbf{1.000} \\
            ClearScope & KAIROS & 0.714 & 1.000 & 0.983 \\
                        & TFLAG    & \textbf{1.000} & 0.800 & \textbf{0.995} \\
            THEIA      & KAIROS & 0.667 & 1.000 & 0.994 \\
                        & TFLAG    & \textbf{1.000} & 0.875 & 0.992 \\
            \bottomrule 
        \end{tabular}
    \end{center}
\vspace{-2.0em}
\end{table}

\begin{figure}
\centering
\includegraphics[width=3in]{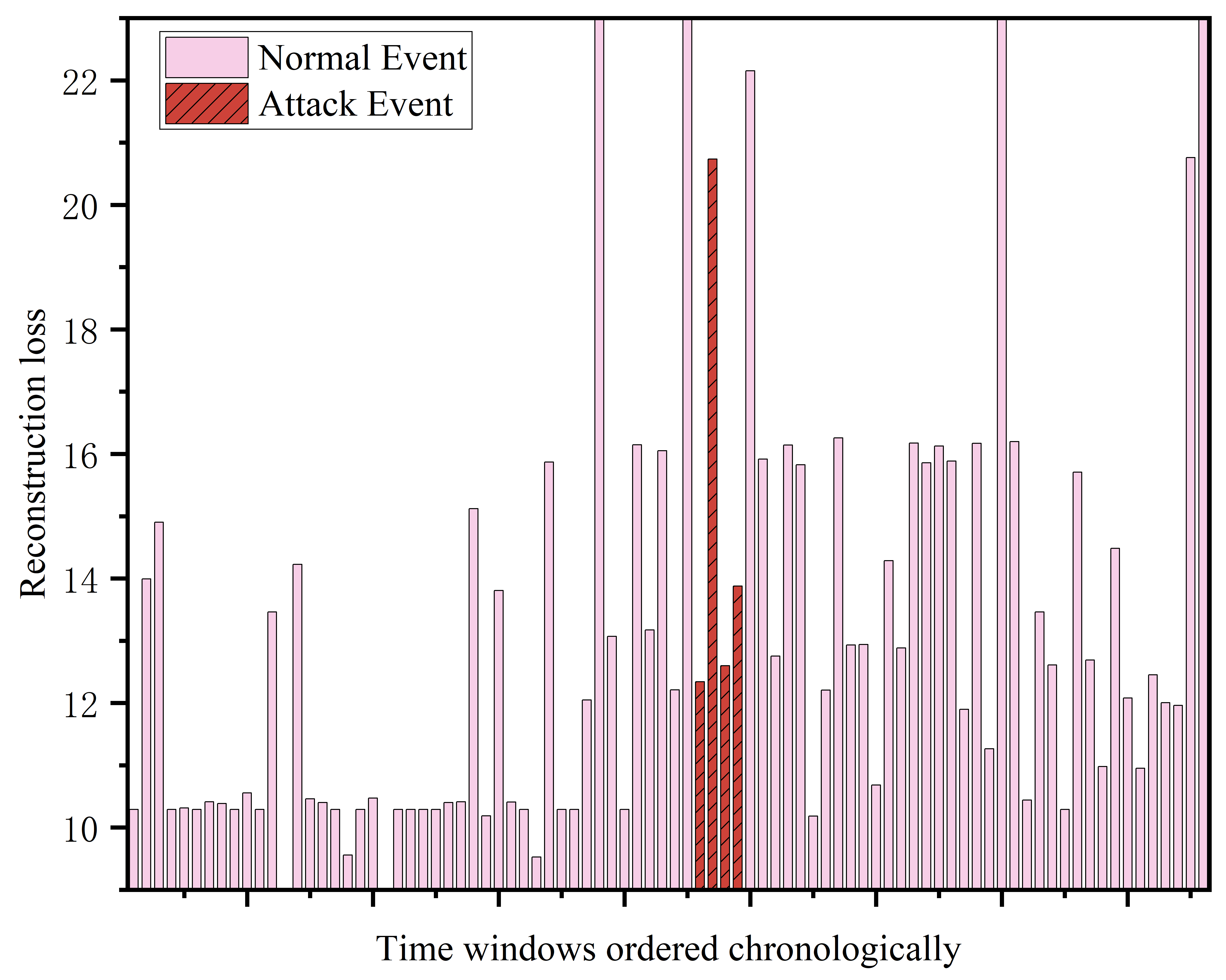}
\caption{The distribution of the highest reconstruction loss events for each time window in the detection results of TFLAG under the no deviation network condition in CADETS-E3 of 2018-04-06.}
\label{nogdn}
\vspace{-1.0em}
\end{figure}

\subsection{Hyperparameter Impact on Performance}

The final detection efficacy of the TFLAG model is significantly influenced by the interaction between its model training and anomaly detection modules, each responsive to hyperparameter settings. This paper examines the impact of varying these parameters on the model's performance in detecting abnormal center windows.

\textbf{Deviation network. }
The deviation network, as one of the core designs of this system, aims to reduce false alarms for attacks deviating from the system’s normal behavior by using self-supervised learning with only normal data in the absence of anomaly labels. As shown in Fig.~\ref{nogdn}, when using the TGAT model for single prediction, the reconstruction loss of most normal events in normal windows exceeds that of true attack events. In contrast, the detection results of TFLAG in Fig.~\ref{loss} demonstrate that the design of the deviation network module is effective in tracing APT attacks in graph-based detection.

\textbf{Neighbor size. }
 In model training, the neighbor size determines the count of neighboring nodes considered before feature aggregation. A size that is too small leads to sparse aggregates, insufficient for capturing comprehensive nodal feature data, which impairs training outcomes and can result in misjudging the abnormal center window, subsequently affecting the accuracy in detecting further anomalies. Conversely, a larger neighbor size considerably increases memory overhead without proportionate improvements in results. The Fig.~\ref{fig:N} illustrates the model's failure to detect the abnormal center window when the neighbor count is configured at either 5 or 10, resulting in zero accuracy. However, optimal performance is observed when the size exceeds 20.

\textbf{Hidden layer dimension. }
This dimension influences the capacity of a node to store aggregated neighbor feature information, encompassing historical, edge, node, and complex temporal data. If the dimension is too minimal, the model inadequately learns from normal data, which compromises edge prediction. Conversely, an overly large hidden layer increases storage and computational demands and might amplify irrelevant features, thus disrupting anomaly detection. The Fig.~\ref{fig:h} demonstrates that setting the dimension to 128 optimizes model efficiency, whereas a dimension of 64 results in false negatives during the detection of attack events, attributed to deviations in the abnormal center window.

\textbf{Edge embedding dimension. }
Edge features, which articulate the relationships between source and target nodes, carry substantial contextual and mutual influence data. Enhancing the dimension of edge embedding allows the model to discern spatial and semantic relations more accurately within a higher-dimensional vector space, which is critical for identifying complex patterns and implicit information. This ability facilitates stronger differentiation among nodes or edges with similar attributes. Nevertheless, exceedingly high dimensions can detract from the model's generalization capabilities, heighten memory usage, and escalate processing requirements, adversely affecting practical performance. As shown in Fig.~\ref{fig:e}, optimal outcomes in terms of accuracy, generalization, and computational efficiency are achieved with an edge embedding dimension set to 16, underscoring the importance of dimension selectivity for optimal model performance.

\textbf{Time window size. }
The chosen size for time windows affects both the number of events per window and the coverage across an entire attack event. For instance, using a 5-minute window for a two-hour attack results in more discrete windows than a 20-minute window. Smaller windows might miss capturing sequential abnormal events due to the prolonged nature of APT attacks, leading to reduced accuracy and recall rates. As illustrated in Fig.~\ref{fig:T}, the analysis of datasets such as CADETS and THEIA indicates that a 15-minute window is optimal, whereas for rapid attack scenarios like those in CLEARSCOPE, a 10-minute window yields the best results.

\subsection{Runtime Performance}
In this experiment, the performance of TFLAG was assessed using benign data collected over two days, with the goal of training the system to accurately identify and predict specific behavioral patterns. Our provenance graph preprocessing technology enabled the successful filtration of 5,372,032 key events. Comparative analysis revealed that the CADETS system typically generates approximately 300 underlying log events per second during normal operation. However, in the offline experiment scenario, the system processed temporal data at a rate of 1,000 items per second, markedly exceeding the data production rate of regular operations.

Additionally, the computational efficiency of TFLAG is influenced by both the optimization of its core algorithm and the number and temporal scope of the processed adjacent nodes. Although TFLAG demonstrates considerable adaptability and high accuracy within dynamically changing real-world scenarios, its processing speed is modest compared to other node-level detection systems in the provenance graph. Nonetheless, by slightly sacrificing speed, it achieves heightened sensitivity in detecting attack events due to differences in detection granularity and the specifics of the final attack detection outcomes.

\section{Conclusion}

We introduce TFLAG, a detection system specifically engineered to identify APT attacks concealed within subtle temporal perturbations. Leveraging the continuous and complex structural properties of provenance graphs, TFLAG combines the dynamic information extraction capabilities of temporal graph models with the anomaly detection capacities of deviation networks in a self-supervised approach, addressing two fundamental challenges in current APT detection methods. It capitalizes on aggregated dynamic historical event features to maintain behavioral continuity, rather than segmentation of APT attacks into discrete snapshots. Moreover, it employs deviation networks to model the anomaly distribution of typical behaviors, thereby effectively distinguishing genuine attack events from sporadic system anomalies. Evaluation results confirm that TFLAG precisely captures low-frequency APT attack behaviors across large datasets, simultaneously filtering out inevitable system irregularities, thus reducing false positives and enhancing detection accuracy.

\bibliographystyle{IEEEtran}
\bibliography{IEEEabrv,ref}

\vspace{-20 mm}
\begin{IEEEbiography}[{\includegraphics[width=1in,height=1.25in,clip,keepaspectratio]{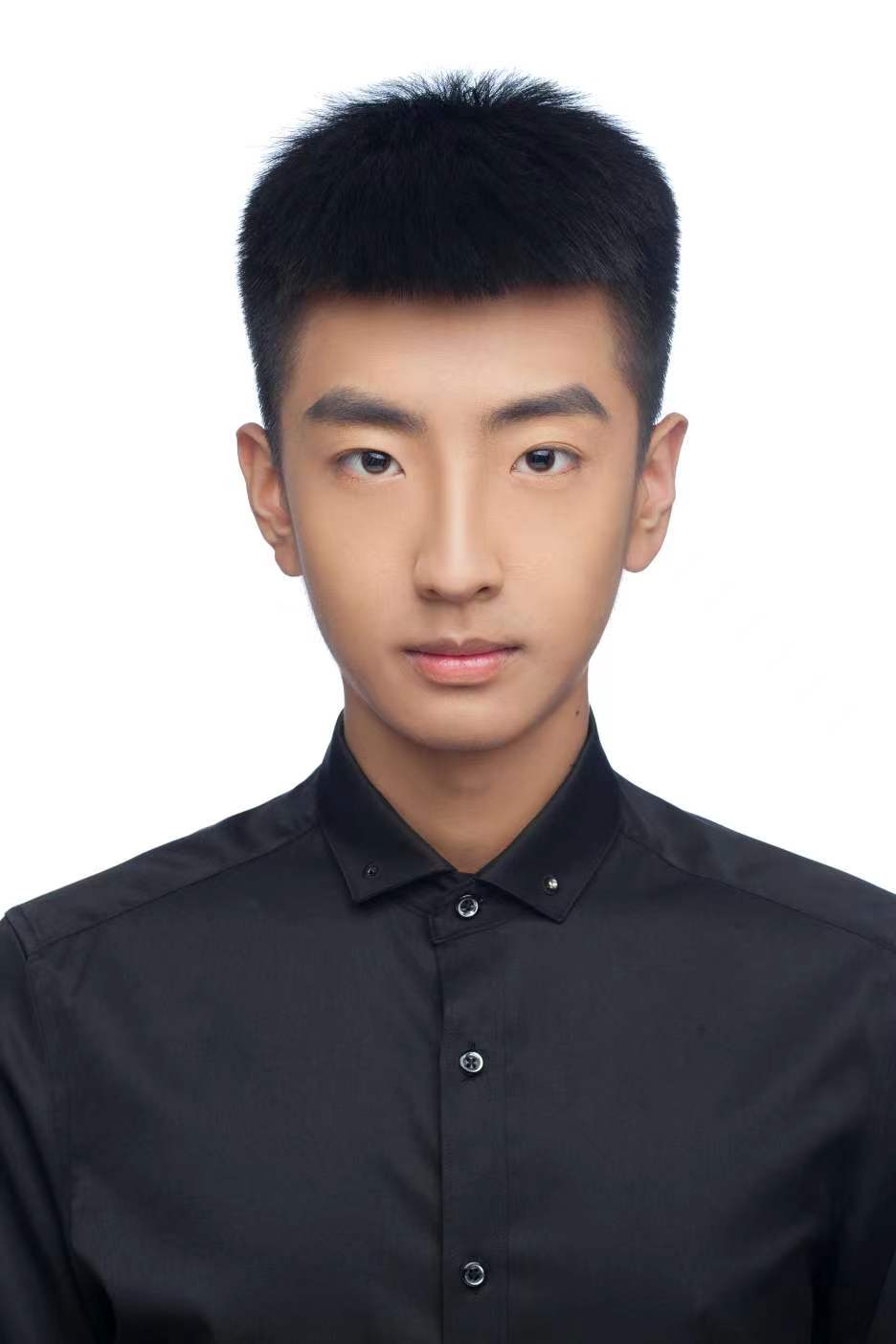}}]{Wenhan Jiang}
received the B.Sc. degree in information security from Hainan University, Haikou, China, in 2023. He is currently pursuing the academic master’s degree in cybersecurity at Harbin Institute of Technology (HIT), Weihai. His current research interests include anomaly detection in provenance graphs and machine learning.
\end{IEEEbiography}

\vspace{-20 mm}
\begin{IEEEbiography}[{\includegraphics[width=1in,height=1.25in,clip,keepaspectratio]{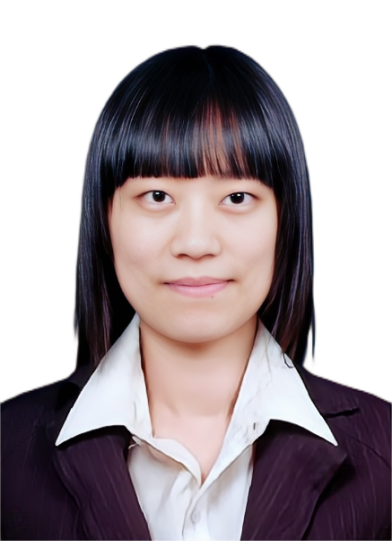}}]{Tingting Chai}
(Member, IEEE) received B.S. degree in Information Security from China University of Mining And Technology in 2012, M.S. degree in Computer Technology from China University of Mining And Technology in 2014 and Ph.D. degree from Institute of Information Science, Beijing Jiaotong University, China in 2020. Currently, she works as an assistant prof. in Faculty of Computing, Harbin Institute of Technology, Harbin, China. Her research interests include Biometrics, Computer Vision and Pattern Recognition.  
\end{IEEEbiography}


\vspace{-20 mm}
\begin{IEEEbiography}[{\includegraphics[width=1in,height=1.25in,clip,keepaspectratio]{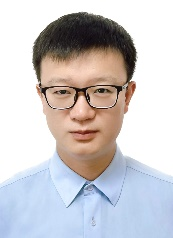}}]{Kai Wang}
(Member, IEEE) is a full Professor with School of Computer Science and Technology (Weihai 264209) and with Faculty of Computing (Harbin 150001), Harbin Institute of Technology, China. His research interests include trustworthy AI and network intrusion detection. He has published more than 40 papers in prestigious international journals, including IEEE TSC, IEEE TITS, ACM TOIT, ACM TIST, etc. He received the Ph.D. degree in Communication and Information Systems from Beijing Jiaotong University, China, in 2014. From 2017 to 2019, he was a postdoc researcher in computer science and technology with Tsinghua University, China. He is a Member of the IEEE and ACM, and a Senior Member of the China Computer Federation (CCF). 
\end{IEEEbiography}

\vspace{-20 mm}
\begin{IEEEbiography}[{\includegraphics[width=1in,height=1.25in,clip,keepaspectratio]{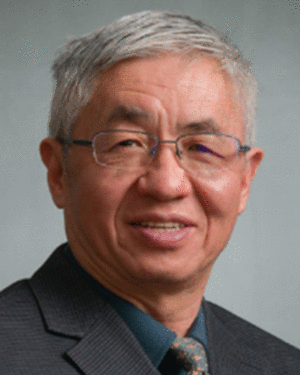}}]{Hongke Zhang}
(Fellow, IEEE) received the Ph.D. degree in communication and information system from the University of Electronic Science and Technology of China, Chengdu, China, in 1992. He is currently a Professor with the School of Electronic and Information Engineering, Beijing Jiaotong University, Beijing, China, where he currently directs the National Engineering Center of China on Mobile Specialized Network. His current research interests include architecture and protocol design for the future Internet and specialized networks. He currently serves as an Associate Editor for the IEEE Transactions on Network and Service Management and IEEE Internet of Things Journal. He is an Academician of China Engineering Academy.
\end{IEEEbiography}

\end{document}